\documentclass[twocolumn,aps,prx,showpacs,amsmath,superscriptaddress,longbibliography,notitlepage]{revtex4-2}
\usepackage{physics,tikz,bm,graphicx,amsmath,amssymb,times,booktabs,multirow}

\usepackage[normalem]{ulem}
\newcommand{\ii}{\mathrm{i}}
\newcommand{\ee}{\mathrm{e}}

\newcommand{\kb}[2]{\left|#1\right\rangle\!\left\langle#2\right|}

\newcommand{\calL}{\mathcal{L}}

\usepackage[colorlinks,linkcolor={blue!75!black},citecolor={blue!75!black},urlcolor={blue!75!black}]{hyperref}

\begin{document}

\title{Universal Dynamical Scaling of Strong-to-Weak Spontaneous Symmetry Breaking in Open Quantum Systems}
\author{Chang Shu}
\affiliation{Department of Physics, University of Michigan, Ann Arbor, MI 48109, United States}
\author{Kai Zhang}
\email{phykai@umich.edu}
\affiliation{Department of Physics, University of Michigan, Ann Arbor, MI 48109, United States}
\author{Zhu-Xi Luo}
\email{zhuxi\_luo@gatech.edu}
\affiliation{School of Physics, Georgia Institute of Technology, Atlanta, GA 30332, United States}
\author{Yizhi You}
\email{y.you@northeastern.edu}
\affiliation{Department of Physics, Northeastern University, Boston, MA, 02115, United States}
\author{Kai Sun}
\email{sunkai@umich.edu}
\affiliation{Department of Physics, University of Michigan, Ann Arbor, MI 48109, United States}

\date{\today}

\begin{abstract}
Strong-to-weak spontaneous symmetry breaking (SWSSB) defines a mixed-state phase of matter--without a pure-state counterpart--in which nonlinear observables such as the Rényi-2 correlator develop long-range order while conventional linear correlations remain short-ranged. Here we study the emergence of SWSSB in one-dimensional open quantum systems governed by Lindbladian evolution, where the transition time diverges with system size and SWSSB appears only asymptotically in the steady state. By tracking the late-time growth of the Rényi-2 correlation length, we uncover a universal dynamical regime controlled purely by the symmetry class of the Lindbladian. 
Contrary to the conventional expectation that late-time dynamics are governed by the low-lying Liouvillian spectrum, we find that the time dependence of the SWSSB transition—exponential versus algebraic—is dictated solely by symmetry, independent of details of the Lindbladian, including whether the Liouvillian spectrum is gapped or gapless. For $\mathbb{Z}_2$-symmetric dynamics, the Rényi-2 correlation length grows exponentially in time--even when the spectrum is gapless--yielding an effective transition time $t_c \propto \operatorname{ln} L$ and enabling rapid preparation of the $\mathbb{Z}_2$ SWSSB steady state. In contrast, U(1)-symmetric dynamics exhibit algebraic scaling, $t_c \propto L^{\alpha}$, with a filling-dependent dynamical exponent: ballistic growth ($\alpha \simeq 1$) at finite filling crosses over to diffusive scaling ($\alpha = 2$) in the zero-filling limit. These results establish symmetry--rather than spectral gap structure--as the controlling principle for SWSSB late-time dynamical scaling, and open a new route to nonequilibrium symmetry breaking in open quantum systems. 
\end{abstract}

\maketitle

\section{Introduction}
A generic mixed quantum state, viewed as an ensemble of pure states, can exhibit both strong and weak symmetries~\cite{Prosen2012,Lieu2020,Albert2014}, where a strong symmetry is respected by every ensemble member state, whereas a weak symmetry holds only at the level of ensemble averaging. 
Without any pure-state counterpart, mixed states can exhibit strong-to-weak spontaneous symmetry breaking (SWSSB), which can be diagnosed by long-range order in the fidelity correlator~\cite{Lessa2025PRXQ} or in the analytically tractable R\'enyi-2 correlator~\cite{FengLei2025PRL,Lee2023}, thereby characterizing a unique mixed-state phases and phase transitions~\cite{Lee2023,Sala2024,Lessa2025PRXQ,FengLei2025PRL,Ma2025,Ellison2025PRX,Guo2025PRB,Shah2025,Luo2025,Ando2026,hauser2026strongtoweaksymmetrybreakingopen}. 
Starting from a strongly symmetric initial state $\rho_0$, it has been shown that SWSSB emerges at a finite decoherence strength under a quantum-channel mapping in two and higher dimensions~\cite{Lessa2025PRXQ}. 
A corresponding continuous-time formulation based on Lindbladian evolution predicts an analogous dynamical transition at a finite evolution time in two and higher dimensional open quantum systems~\cite{Chen2025,xue2025,zhao2025, Sa2025, Gu2025}. 
The critical behavior near the SWSSB transition can be understood by an exact mapping to a well-studied equilibrium phase transition~\cite{Lessa2025PRXQ,Sala2024,Guo2025,Gu2025,Sa2025,Kuno2024,zhao2025,Zhang2025,Zerba2025DipoleConservingSWSSB}. 
In the vicinity of SWSSB, universal dynamical critical exponents can be defined~\cite{Kuno2025}, analogous to those characterizing equilibrium phase transitions. 

Although SWSSB has recently attracted significant attention, its dynamical transition in one dimension has so far received comparatively little attention due to the absence of a well-defined transition parameter. 
In one dimension, the onset time of SWSSB diverges with system size, and the critical time is pushed to infinity in the thermodynamic limit. 
However, this regime is of direct relevance for real-time Liouvillian dynamics in realistic settings. 
In particular, even if the dynamical transition is formally shifted to infinite time in the thermodynamic limit, the R\'enyi-2 correlator may become nonlocal on an exponentially short timescale for any finite system size. In this sense, finite systems effectively enter the SWSSB phase rapidly, despite the absence of a true transition at finite time in the infinite-size limit.
Moreover, the asymptotic long-time scaling behavior of the R\'enyi-2 correlator provides a natural classification of mixed-state phases and their associated non-equilibrium dynamical processes. 
This makes the investigation of its late-time dynamical scaling behavior in SWSSB transition essential in open quantum Lindbladian dynamics. 

For Lindbladian dynamics, it is conventionally expected that the late-time behavior is controlled by the low-lying modes: a gapped spectrum leads to exponential relaxation toward the steady state, with a rate set by the spectral gap, whereas a gapless Liouvillian gives rise to slow, typically algebraic, dynamics with dynamical exponents determined by these modes~\cite{Marko2015,CaiZi2013PRL,Haga2021,Mori2023}.
However, the SWSSB phenomenon is also closely tied to the symmetry structure of the Lindbladian dynamics. In particular, the emergence of SWSSB reflects the spatial spreading of symmetry-charged operators and the ability of local measurements to access the associated symmetry charge. 
These considerations suggest that symmetry itself may play a crucial role in governing the dynamical scaling of the SWSSB transition. 
This naturally raises two key questions: Is the late-time dynamical scaling of SWSSB primarily determined by the Liouvillian spectral structure --- such as the presence or absence of a gap --- or by the underlying symmetry class of the Lindbladian dynamics? Furthermore, does it exhibit universal dynamical scaling with system size? 

In this paper, we find that, in contrast to the conventional wisdom that late-time dynamics are governed by low-lying spectral properties, the symmetry itself dictates the rate of strong-to-weak symmetry breaking in the long-time limit of Lindbladian evolution. 
Specifically, in 1D, when the Liouvillian possesses a discrete symmetry (e.g., $\mathbb{Z}_2$), the R\'enyi-2 correlator spreads exponentially fast regardless of whether the Liouvillian spectrum is gapped or gapless. 
This implies that the SWSSB transition occurs at a critical evolution time scaling as $t_c \propto \ln L$, with $L$ the system size. 
By contrast, for a Liouvillian with a continuous symmetry (e.g., $U(1)$), the spectrum is necessarily gapless and captured by an effective hydrodynamic description~\cite{Lee2023PRL,WangZhong2025PRB}.
In this symmetry case, we find that the late-time dynamics of R\'enyi-2 correlator follows a power-law scaling, with the critical time obeying $t_c \propto L^{\alpha}$. 
For a $U(1)$-symmetric Liouvillian, the value of this exponent $\alpha$ depends on the filling: at the limit of zero or full filling, the dynamics is diffusive with $\alpha = 2$, whereas at any finite filling it becomes near ballistic, yielding $\alpha \simeq 1$. 
These results are in full agreement with our analytical and numerical simulations using two representative quantum spin-chain examples.

The rest of the paper is organized as follows: In Sec.~\ref{sec:infini}, we introduce the notion of infinite-time SWSSB in Lindbladian dynamics.
In Sec.~\ref{sec:Z2}, we show that in $\mathbb{Z}_2$ symmetric models the Rényi-2 correlation length grows exponentially in time even when the spectrum is gapless.
We first show in Sec.~\ref{sec:STNSSB} with a model with closed dynamical sector.
The exponential growth of Rényi-2 correlator reflects an exponentially fast convergence into that sector, while the gapless Lindbladian spectrum controls the slow dynamics within the sector itself.
We observe a broad time window $\ln L\lesssim t\lesssim L^2$ within which the system exhibits SWSSB order. 
We further show in Sec.~\ref{sec:robust} that this ultra-fast scaling persists in a deformed gapless $\mathbb{Z}_2$ model where the special closed dynamical sectors are destroyed, demonstrating that it is a robust consequence of discrete symmetry rather than a fine-tuned sector structure.
Finally in Sec.~\ref{sec:info}, we show the same dynamical separation is also visible via information-theoretic quantities. 
In Sec.~\ref{sec:U1}, we analyze U(1) symmetric models, where the corresponding growth is algebraic and filling dependent. We conclude in Sec.~\ref{sec:outlook} with a discussion and outlook.

\section{Infinite-time SWSSB transition}\label{sec:infini}

Prior to analyzing the late-time dynamical scaling behavior, we first clarify the definition of SWSSB in Lindbladian dynamics. 
Although both the fidelity and the R\'enyi-2 correlator can signal the onset of SWSSB~\cite{Lessa2025PRXQ}, throughout this work we employ the R\'enyi-2 correlator as a diagnostic due to its direct computability~\cite{FengLei2025PRL}. 
Consider a strong symmetry $g\in G$ with unitary representation $U_g$. 
Let $O_x$ denote a local charge operator that transforms nontrivially under the symmetry.
The R\'enyi-2 correlator is defined as
\begin{equation}\label{eq_Renyi2}
    R_2(r,t)=\frac{\Tr[\rho(t) O_x^\dagger O_y\rho(t) O_y^\dagger O_x]}{\Tr[\rho(t)^2]},
\end{equation}
where we assume translation invariance and define $r:=|x-y|$.  
Without loss of generality, we consider an initial state $\rho_0$ that respects the strong symmetry $U_g$, i.e., $U_g\rho_0 = e^{i\theta_g}\rho_0$, and exhibits short-range R\'enyi-2 correlations, i.e., $\lim_{r\rightarrow \infty} R_2(r,t=0) = 0$. 
The state evolves under the Lindblad master equation
\begin{equation}\label{eq_masterequation}
    \frac{d\rho}{dt} = -i[H,\rho] + \sum_{\mu} L_{\mu} \rho L_{\mu}^{\dagger} -\frac{1}{2}\{L_{\mu}^\dagger L_\mu, \rho\} \equiv \mathcal{L}\rho,
\end{equation}
where the Liouvillian preserves the strong symmetry, $[H,U_g]=0$ and $[L_{\mu},U_g]=0$ for all $\mu$. 
In the long-time limit, the system approaches a steady state $\rho_{s}$ that remains in the same symmetry sector as the initial state. 
Suppose that the steady state exhibits long-range R\'enyi-2 correlations, $\lim_{r\rightarrow \infty} R_2(r,\infty) = \mathcal{O}(1)$, then SWSSB must occur during the dissipative evolution. 

Two distinct dynamical SWSSB transition scenarios arise. 
(i)~SWSSB occurs at a finite time. 
There exists a critical time  $t_c < \infty$ such that for $t < t_c$ the correlations remain short-ranged, satisfying $\lim_{r\to\infty} R_2(r,t<t_c)=0$, whereas for  $t > t_c$ the correlator develops long-range order, $\lim_{r\to\infty} R_2(r,t>t_c)=\mathcal{O}(1)$. 
Examples of finite-time SWSSB have been discussed in Refs.~\cite{Sala2024, Zhang2025, zhao2025, Sa2025}. 
(ii)~SWSSB occurs at the infinite time $t_c=\infty$. 
The critical time is pushed to infinity $t_c=\infty$. 
In this case, for any finite time the R\'enyi-2 correlation remains short-ranged, 
\begin{equation}\label{eq:avoid_finite_tc}
    \lim_{r\to\infty}R_2(r,t<\infty)=0,
\end{equation}
while long-range order emerges only asymptotically in the steady state 
\begin{equation}\label{eq:longrangesteadystate}
    \lim_{r\to\infty}\lim_{t\to\infty}R_2(r,t)=\mathcal{O}(1).
\end{equation}
The noncommutativity between the limits $r\to\infty$ and $t\to\infty$ characterizes the infinite-time SWSSB transition in Lindbladian dynamics.
This definition of infinite-time SWSSB will serve as the starting point for our analysis of late-time dynamical scaling below. 

For the infinite-time SWSSB transition, we consider a 1D open quantum spin chain with only jump operators $L_{j}=\sqrt{\gamma} Z_j Z_{j+1}$. 
In this case, the R\'enyi-2 correlator can be solved exactly and takes the form $R_2(r,t)=A\ee^{-r/\xi(t)}$ with a time-dependent correlation length $\xi(t)=\ee^{4\gamma t}$~[see Appendix.~\ref{A:Z2Gap} for derivations]. 
We can verify that this satisfies both conditions in Eq.~\eqref{eq:avoid_finite_tc} and Eq.~\eqref{eq:longrangesteadystate}, signaling a infinite-time SWSSB transition. 
More generally, for systems exhibiting an infinite-time SWSSB transition, the correlation length must diverge asymptotically, $\xi(t\to\infty)\to\infty$. 
This naturally raises two central questions: (i) how does the divergence of the correlation length scale with system size, and (ii) what determines the corresponding dynamical scaling law?
In the $ZZ$-decoherence model discussed above, the correlation length grows exponentially, $\xi(t)\sim e^{4\gamma t}$, indicating ultra-fast spreading of R\'enyi-2 correlation. 
In this specific model, the exponential growth is directly related to the finite Liouvillian gap $\Delta=2\gamma$.
However, when the Liouvillian spectrum becomes gapless in a $\mathbb{Z}_2$ symmetric model, it is unclear whether such exponentially rapid spreading remains robust since a well-defined Liouvillian gap is absent then.
One might instead expect that, for systems exhibiting infinite-time SWSSB but possessing a gapless Liouvillian spectrum, the growth of $\xi(t)$ becomes qualitatively slower, for example sub-exponential.
As we demonstrate below, however, this intuition is incomplete: the symmetry structure, rather than the Liouvillian gap, ultimately governs the late-time scaling of the R\'enyi-2 correlation. 

\section{Ultra-fast scrambling in gapless \texorpdfstring{$\mathbb{Z}_2$}{Z2}-symmetric models}\label{sec:Z2}
\subsection{Exponentially fast SWSSB in gapless $\mathbb{Z}_2$ dynamics with closed dynamical sectors}\label{sec:STNSSB}
Here, we present a one-dimensional $\mathbb{Z}_2$ strongly symmetric Lindbladian model whose spectrum is gapless and the steady state exhibits SWSSB. 
One might expect that gaplessness necessarily leads to slow, power-law spreading of correlations.
Instead, we show that the Rényi-2 correlation diagnosing SWSSB spread exponentially fast in time.
This behavior is closely tied to the presence of a strong global $\mathbb{Z}_2$ symmetry, which leads to an exponentially fast decay of off-diagonal coherences, while the space spanned by these slow modes becomes dynamically closed and does not contribute to the emergence of SWSSB. 

\begin{figure}[t]
  \centering
  \includegraphics[width=\linewidth]{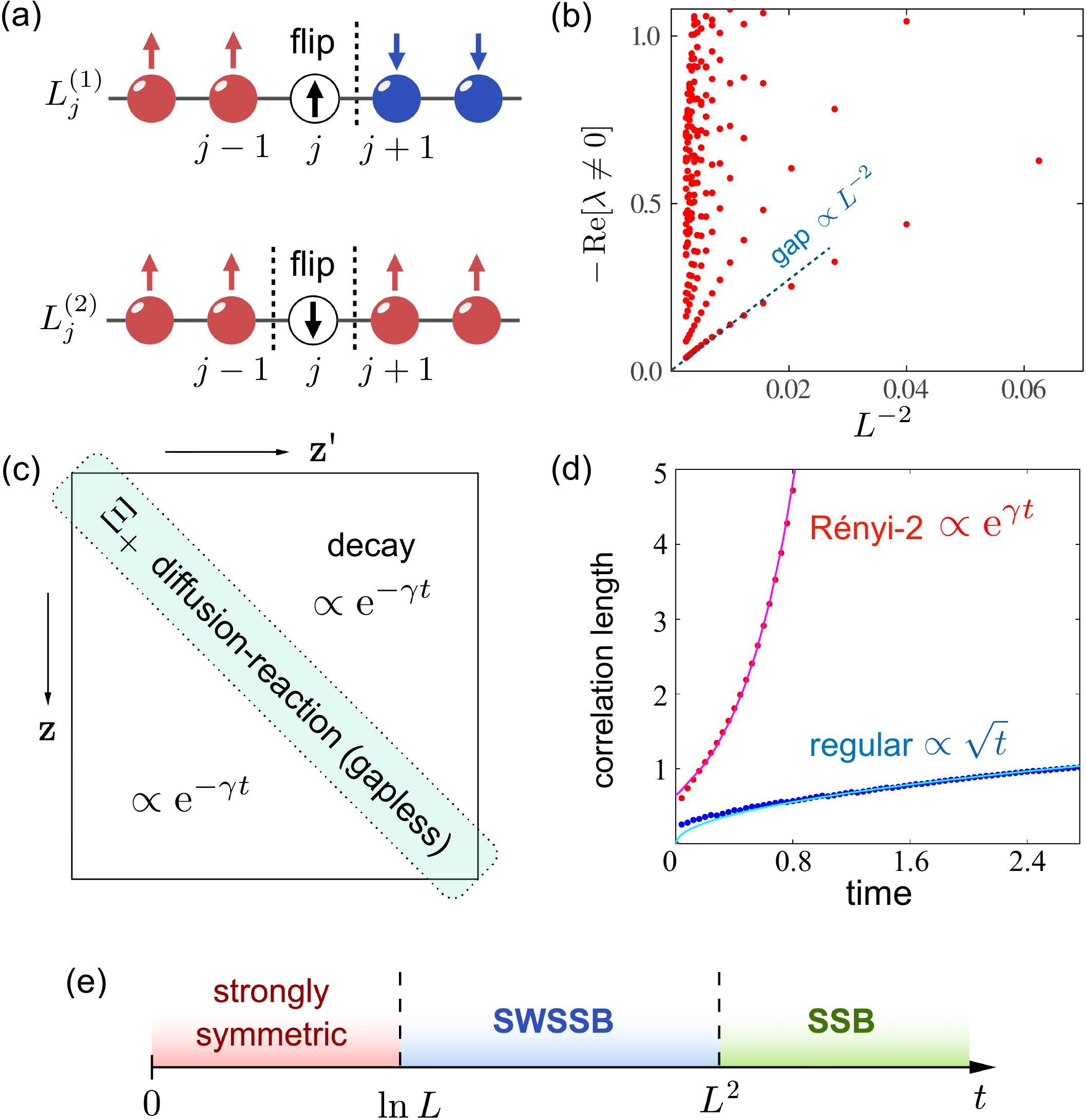}
  \caption{
  Gapless $\mathbb{Z}_2$ symmetric Lindbladian model [$\gamma_1=\gamma_2=1.0$].
    (a) $L^{(1)}$ moves the $z$-domain wall by one unit length, while $L^{(2)}$ annihilates a pair of adjacent domain walls.
    The dashed lines represent domain walls.
    (b) Finite-size scaling of the first 50 eigenvalues (except for zeros) of the Liouvillian restricted to the diagonal sector $\Xi$ for system size up to $L=20$. The gap size scales as $\delta\propto L^{-2}$ [blue dashed line].
    (c) shows the structure of the density matrix in the $z$ basis. The slow gapless mode is contributed by the sector with $\mathbf{z}=\pm\mathbf{z}'$.
    (d) TEBD simulation of a $L=150$ chain shows that the Rényi-2 correlation length (red dots) shows exponential growth (magenta fit) while the regular correlation length (blue dots) shows square-root growth (cyan fit).
    (e) For a finite system of size $L$, three distinct regimes can be observed in the dynamical evolution, where the onset of SWSSB happens in time window $\ln L\lesssim t \lesssim L^2$.
  }
  \label{fig:1}
\end{figure}

Consider a qubit chain of length $L$ with Pauli operators $X_j,Z_j$ locally acting on site $j$.
The dissipative dynamics is generated by two families of local jump operators,
\begin{equation}\label{eq_gaplessZ2}
  \begin{aligned}
    L^{(1)}_{j} &= \sqrt{\gamma_1}\, X_j\, P_{j-1,j+1},\\
    L^{(2)}_{j} &= \sqrt{\gamma_2}\, X_j\, P_{j,j-1}P_{j,j+1},
  \end{aligned}
\end{equation}
where $\gamma_{1,2}>0$ are dissipation rates and $P_{i,j}:=(I-Z_i Z_j)/2$ projects onto configurations with $z_i z_j=-1$, i.e., spins at sites $i$ and $j$ anti-parallel along the $z$ axis.
To clarify the physical meaning, it is convenient to work in the $z$ basis and describe configurations in terms of domain walls.
The jump operator $L^{(1)}_j$ acts only when $z_{j-1}$ and $z_{j+1}$ are anti-parallel.
Then there is precisely one domain wall adjacent to site $j$: flipping the spin at $j$ moves this domain wall by one lattice spacing [Fig.~\ref{fig:1}(a)].
Thus $L^{(1)}$ implements diffusion of domain walls.
The second jump operator $L^{(2)}_j$ requires the spin at site $j$ is opposite to both neighbors, corresponding to two adjacent domain walls sandwiching site $j$.
Applying $X_j$ flips $z_j$, aligning it with both neighbors and removing both domain walls at once [Fig.~\ref{fig:1}(a)].
Therefore, $L^{(2)}$ realizes pair annihilation of domain walls.
It directly follows that the dark states (or decoherence-free subspace~\cite{Lidar1998,Kraus2008,Lidar2014ReviewDFS}) regarding these two types of Lindbladian jump operators are spanned by $\ket{U}:=\otimes_j\ket{\uparrow}_j$ and $\ket{D}:=\otimes_j\ket{\downarrow}_j$---the two ferromagnetic (FM) product states with no domain walls in the $z$ basis. 
The Liouvillian spectrum is gapless, as shown in Fig.~\ref{fig:1}(b).
The finite-size scaling is obtained via exact diagonalization of the full Liouvillian, showing that the Liouvillian gap scales as $\delta\propto L^{-2}$. 

We now examine the symmetry breaking patterns in the the dissipative dynamics governed by Lindblad master equation in Eq.~\eqref{eq_masterequation}. 
This Lindbladian has strong $\mathbb{Z}_2$ symmetry $X=\prod_j X_j$. 
Therefore, the parity of the density matrix, $X\rho_{\pm}=\pm\rho_{\pm}$, will be preserved during the Lindbladian time evolution and determined by the initial state.
Denoting $\ket{\pm}=(\ket{U}\pm\ket{D})/\sqrt{2}$, the doubly degenerate steady states of the Lindbladian are $\rho^{\pm}_{\text{ss}}=\kb{\pm}{\pm}$, where the sign denotes the $X$-parity. 
Since these steady states are precisely the symmetric and antisymmetric GHZ states, the strong Ising symmetry has been broken down to trivial, which can be seen from explicitly computing the conventional two-point correlator $C(|i-j|, t):=\Tr[\rho(t)Z_i Z_j]$. However, below we will show that there is separation of time scales between the strong-to-weak and weak-to-trivial symmetry breakings. 

We start with a strongly symmetric initial state $\ket{\psi_0}=\otimes_j\ket{\rightarrow}_j$, for which the R\'enyi-2 correlation is short-ranged, $R_2(|i-j|,0)=\delta_{i,j}$. To detect strong symmetry breaking, we examine the following Rényi-2 correlator based on the local charged operator $Z_j$,
\begin{equation}\label{eq:Renyi2}
  R_2(|i-j|,t)=\frac{\Tr[\rho(t)Z_i Z_j\rho(t)Z_j Z_i]}{\Tr[\rho(t)^2]}.
\end{equation}
It is straightforward to verify that $R_2(|i-j|,\infty)\equiv 1$, i.e., the steady state of the Lindbladian has long-range order in terms of Rényi-2 correlation.
Moreover, the Rényi-1 correlator
\begin{equation}\label{eq:fidelity}
    R_1(|i-j|,t)=\Tr[\sqrt{\rho(t)}Z_i Z_j\sqrt{\rho(t)} Z_j Z_i]
\end{equation}
is also order long-ranged at $t\to\infty$, signaling strong symmetry breaking in the steady state in the fidelity sense~\cite{Weinstein2025,Liu2025}.

At generic times away from phase transitions, taking $r:=|i-j|$, the Rényi-2 correlation can be written as $R_2(r,t)=A\ee^{-r/\xi(t)}$, where $\xi$ represents the correlation length. 
Under the Lindblad time evolution Eq.~\eqref{eq_masterequation}, $R_2(r,t)$ spreads throughout the system and approaches a spatially nonlocal profile. 
As shown in Fig.~\ref{fig:1} (d), the R\'enyi-2 correlation length growth exponentially in time, whereas the conventional correlation length grows only diffusively, $\xi_{\text{reg}}(t)\sim \sqrt{t}$.
For a finite system of size $L$, this means the nonlinear correlator spans across the system already at $t\sim\ln L$, while the conventional correlator only reaches the system size at $t\sim L^2$.
Hence, there exists a broad time window, $\ln L\lesssim t\lesssim L^2$, in which the system exhibit long-range order in R\'enyi-2 correlator while the conventional correlator stays short-ranged [Fig.~\ref{fig:1} (e)].

Now we account for the observed phenomena by revealing the existence of a slow closed dynamical sector decoupled from the SWSSB physics.
In general, the density matrix in $z$ basis can be expanded by
\begin{equation}\label{eq:rhoinz}
    \rho(t)=\sum_{\mathbf{z},\mathbf{z}'}p_{\mathbf{z},\mathbf{z}'}(t)\ket{\mathbf{z}}\bra{\mathbf{z}'}
\end{equation}
where $\mathbf{z}=(z_1,...,z_n)$ denotes a spin-$z$ configuration.
Given that the initial state $\rho_0=2^{-L}\sum_{\mathbf{z} \mathbf{z}'}\kb{\mathbf{z}}{\mathbf{z}'}$ and the evolution is purely dissipative without coherent pieces, the coefficient $p_{\mathbf{z},\mathbf{z}'}=p_{\mathbf{z}',\mathbf{z}}$ remain real.
Denote the action of jump operators as $L_j^{(\alpha)}\ket{\mathbf{z}}=q_j^{(\alpha)}(\mathbf{z})|\mathbf{z}^{(j)}\rangle$ where $\mathbf{z}^{(j)}$ represents the configuration obtained from flipping the spin at $j$-site.
Due to $\mathbb{Z}_2$ symmetry, we have $q_j^\alpha(\mathbf{z})=q_j^\alpha(-\mathbf{z})$.
From the Lindblad master equation, we get
\begin{equation}\label{eq:pODEqq}
\begin{aligned}
    \frac{\dd p_{\mathbf{z},\mathbf{z}^{\prime}}}{\dd t}=\sum_{j=1}^{L}\sum_{\alpha=1,2}&q_j^\alpha({\mathbf{z}^{(j)}})q^{\alpha}_j(\mathbf{z}^{\prime(j)})p_{\mathbf{z}^{(j)},\mathbf{z}^{\prime(j)}}\\-&\sum_{j=1}^{L}\sum_{\alpha=1,2}\frac{q_j^\alpha(\mathbf{z})^2{+}q_j^\alpha(\mathbf{z}')^2}{2}{p_{\mathbf{z},\mathbf{z'}}}.
\end{aligned}
\end{equation}
Now, we show the dynamical sectors $\Xi^{\pm}:=\{\kb{\mathbf{z}}{\mathbf{z}'}:\mathbf{z}=\pm\mathbf{z}'\}$ are responsible for the gapless spectrum.
Notice that these sectors $\Xi^{+},\Xi^{-}$ are both dynamically closed under the Lindbladian evolution.
When $\mathbf{z}=\pm\mathbf{z}'$, we denote $P_{\pm}(\mathbf{z}):=p_{\mathbf{z},\pm\mathbf{z}}$ and get
\begin{equation}\label{eq:MarkovChain}
    \frac{\dd P_{\pm}(\mathbf{z})}{\dd t}=\sum_{j=1}^L[W_{\mathbf{z}^{(j)}\to\mathbf{z}}\,P_{\pm}(\mathbf{z}^{(j)})-W_{\mathbf{z}\to\mathbf{z}^{(j)}}\,P_{\pm}(\mathbf{z})]
\end{equation}
where $W_{\mathbf{z}\to\mathbf{z}^{(j)}}=\sum_{\alpha=1,2}[q_j^\alpha(\mathbf{z})]^2$ is the transition rate from $\mathbf{z}$ to $\mathbf{z}'$ configuration.
This is a standard continuous-time Markov chain equation~\cite{Norris1997MarkovChains,Breuer2002OpenQuantum}.
The dynamics within $\Xi^{\pm}$ is the classical diffusion-reaction process~\cite{Glauber1963TimeDependentIsing,LUSHNIKOV1987135}.
When confining the action of the Liouvillian inside the section $\Xi$, we obtain gapless spectrum [see Appendix.~\ref{A:Z2Gapless} for proof].
At late time, what remains is a dilute domain wall gas where its density tends to zero.
For any pair of the domain walls separated by distance $\sim L$, it takes $\Delta t\sim L^2$ for them to annihilate each other due to the diffusive nature of the diffusion-reaction process described by Eq.~\eqref{eq:MarkovChain}.
Therefore, the energy scale for this late-time dynamics is of the order $\sim L^{-2}$, consistent with the observation in Fig.~\ref{fig:1}(b). Notably, the gapless Liouvillian spectrum and the resulting long mixing time can be understood as consequences of an emergent strong \(U(1)\) symmetry in the low-energy sector of the Liouvillian. We elaborate on this point in Appendix.~\ref{app:U1}.

Despite the gapless nature of the $\Xi^{\pm}$ sectors, we find that the Rényi-2 correlator spreading is exponentially fast, contrary to the simple intuition that gaplessness usually leads to slow late-time dynamics. 
As shown in Fig.~\ref{fig:1} (d), we numerically simulated a qubit chain of length $L=150$ with dissipation strength $\gamma_1=\gamma_2=1.0$, using time-evolving block decimation (TEBD) methods~\cite{Zwolak2004,Vidal2003,ITensor}
[Details about the implementing Lindbladian dynamics using TEBD can be found in Appendix.~\ref{A:TEBD}].
The result shows that, at late time, the $\ln \xi$ in $R_2(r,t)$ scales linearly with $t$, signaling the exponentially fast propagation [Fig.~\ref{fig:1} (d)].
The reason becomes clear when we consider the R\'enyi-2 correlator for a generic state in the sector $\Xi^+$, written as $\rho_{\text{d}}=\sum_{\mathbf{z}}c_{\mathbf{z}}\kb{\mathbf{z}}{\mathbf{z}}$, where the coefficients $c_{\mathbf{z}}\in(0,1)$.
Plugging in Eq.~\eqref{eq:Renyi2}, we get $R_2\equiv 1$, fully independent of the spatial coordinates.
This is also the case for the $\mathbb{Z}_2$-symmetry-related sector $\Xi^-$.
This shows that the slow dynamics controlled by the gapless spectrum \textit{does not} contribute to the emergence of SWSSB. 
Therefore, the initial localized R\'enyi-2 correlation is fully determined by the states outside the $\Xi^{\pm}$ sector. 
In fact, the weights outside the $\Xi^{\pm}$ sectors experience exponential decay.
We can rearrange Eq.~\eqref{eq:pODEqq} as
\begin{equation}\label{eq:attrition}
    \frac{\dd p_{\mathbf{z},\mathbf{z}'}}{\dd t}{=}\sum_{j=1}^L [W^j_{\text{in}}p_{\mathbf{z}^{(j)},\mathbf{z}^{\prime (j)}}{-}W_{\text{out}}^j p_{\mathbf{z},\mathbf{z}'}]-\sum_{j=1}^L\kappa^j p_{\mathbf{z},\mathbf{z}'}
\end{equation}
where the weights $W_{\text{out}}^j=\sum_{\alpha}q_j^{\alpha}(\mathbf{z}^{(j)})q_j^\alpha(\mathbf{z}^{\prime(j)})$ and $W_{\text{in}}^j=\sum_{\alpha}q_j^{\alpha}(\mathbf{z})q_j^\alpha(\mathbf{z})$ in the first term still maintains Markovian dynamics.
However, the addition term with $\kappa^j=\sum_{\alpha}(q_j^\alpha(\mathbf{z})-q_j^\alpha(\mathbf{z}'))^2/2$ is a pure ``sink" term contributing to total attrition.
When $\mathbf{z}=\pm \mathbf{z}'$, we have $\kappa^j\equiv 0$, reducing Eq.~\eqref{eq:attrition} to the diffusion-reaction Markov process described by Eq.~\eqref{eq:MarkovChain}.
When we are not in the $\Xi^{\pm}$ sector, the domain wall configuration in $\mathbf{z}$ and $\mathbf{z}'$ mismatch and we have $\sum_j \kappa^j\geq\min\{\gamma_1,\gamma_2\}/2>0$.
Consequently, generic off-diagonal matrix elements have a finite decay rate, causing the Rényi-2 correlation to spread exponentially fast despite a gapless Liouvillian spectrum. 
In addition the fidelity correlator~\cite{Lessa2025PRXQ} and the Rényi-1 correlator $R_1(|i-j|,t)$ in Eq.~\eqref{eq:fidelity} is also expected to exhibit exponentially fast spreading dynamics. 
This follows because, for any mixed-state component lying in the closed dynamical sectors $\Xi^\pm$, its Rényi-1/fidelity correlator is also featureless in space, hence not contributing to its spreading dynamcis.

\subsection{Robustness of exponential scaling in $\mathbb{Z}_2$ models beyond closed dynamical sectors}\label{sec:robust}

In the preceding model, although the Liouvillian is gapless and therefore the relaxation toward the steady state is slow, the spreading of the Rényi-2 correlation is decoupled from these slow modes and can still exhibit fast growth.
A immediate following question is whether the exponential scaling is tied to the special closed-sector structure $\Xi^{\pm}$ of that example discussed near Eq.~\eqref{eq:MarkovChain}, or whether it reflects a more general consequence of discrete symmetry.
To address this, we consider a deformed $\mathbb{Z}_2$-symmetric Lindbladian
\begin{equation}\label{eq_gaplessZ2_2}
  \begin{aligned}
    \tilde{L}^{(1)}_{j} &= \sqrt{\gamma_1}\, (X_j P_{j-1,j+1}+X_{j+1} P_{j,j+2}),\\
    \tilde{L}^{(2)}_{j} &= \sqrt{\gamma_2}\, X_j P_{j,j-1}P_{j,j+1}.
  \end{aligned}
\end{equation}
This deformation preserves the original global strong $\mathbb{Z}_2$ symmetry $X$ and the GHZ steady states, but qualitatively changes the internal structure of the dissipative dynamics.  
In particular, the modified $\tilde{L}^{(1)}_{j}$ implements a coherent superposition of two neighboring domain-wall moves, so that the sectors $\Xi^{\pm}$ are no longer dynamically closed. 
As a result, the Markov reduction used in Sec. III does not apply here, and the exponential growth of the Rényi-2 correlator is no longer guaranteed by the previous closed-sector argument alone.\\

\begin{figure}[t]
    \centering
    \includegraphics[width=\linewidth]{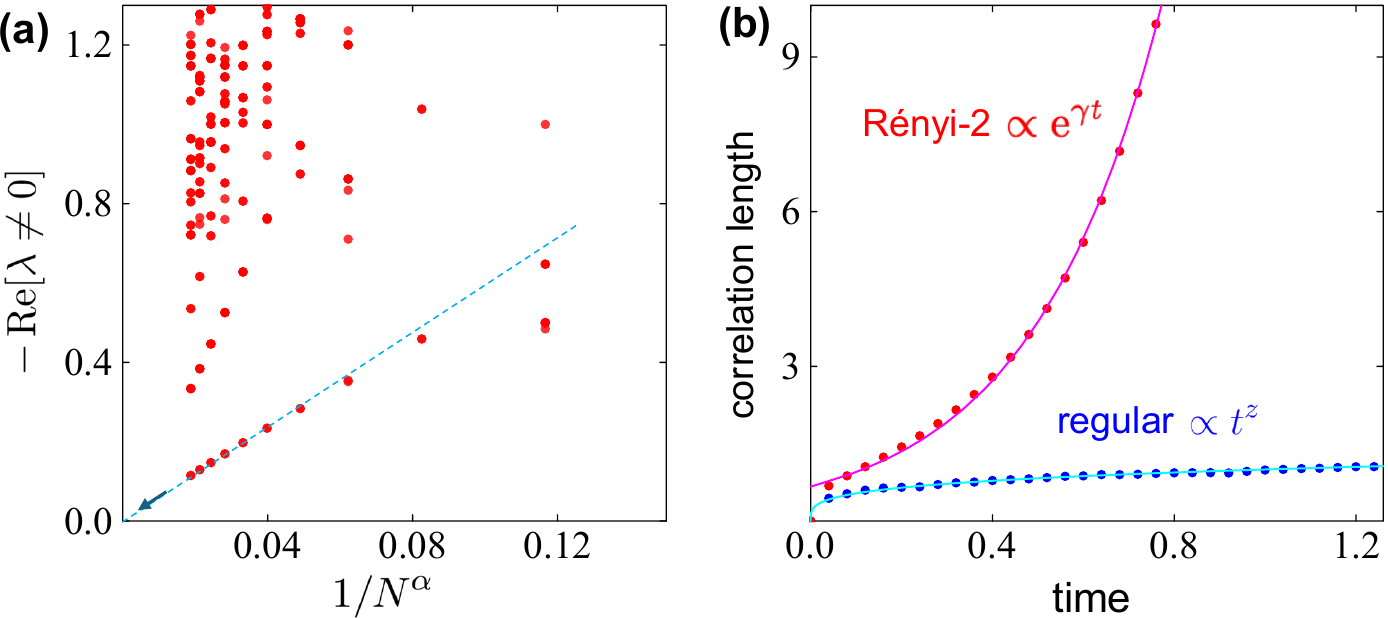}
    \caption{Modified gapless $\mathbb{Z}_2$-symmetric model with $\tilde L^{(1)}_j$ and $\tilde L^{(2)}_j$ given by Eq.~\eqref{eq_gaplessZ2_2} and $\gamma_1=\gamma_2=1$. 
    (a) Finite-size scaling of nonzero Liouvillian eigenvalues from ED for $N=4,5,\cdots,13$, showing a vanishing gap size $\Delta\propto N^{-\alpha}$ with $\alpha = 1.55$.
    (b) Correlation lengths from TEBD: the Rényi-2 correlation length (red dots) grows exponentially, $\xi_{R_2}\sim \ee^{\gamma t}$, while the regular correlation length (blue dots) grows algebraically, $\xi_{\mathrm{reg}}\sim t^{z}$ with $z\simeq 0.27$.}
    \label{fig:2}
\end{figure}

Nevertheless, our numerical results suggest the deformed model remains gapless and still exhibits ultra-fast spreading of the Rényi-2 correlator.
Since the steady state contains no domain walls, the late-time dynamics is still governed by a dilute domain-wall picture.
The key difference is that the deformation in $\tilde{L}_j^{(1)}$ qualitatively modifies the dispersion relation of the domain-wall motion: instead of the diffusive scaling $\Delta t\propto L^2$ in the original model, the characteristic timescale now scales with a modified dynamical exponent $\Delta t\propto L^\alpha$.
As a result, the Liouvillian spectrum remains gapless, but with a modified finite-size gap scaling $\sim L^{-\alpha}$.
Exact diagonalization of the full Liouvillian [Fig.~\ref{fig:2} (a)] for $\gamma_1=\gamma_2=1.0$ and system size $N=4,5,\cdots,13$ shows that the slowest nonzero eigenvalues $\lambda_1$ collapse under the scaling $-\Re \lambda_1\propto N^{-\alpha}$ with fitted value $\alpha=1.55$.
This provides direct numerical evidence that the deformed model remains gapless. In Appendix~\ref{app:U1}, we also provide a complementary theoretical argument showing that the Liouvillian spectrum is necessarily gapless whenever the Markovian dynamics does not increase the domain-wall number, as this condition gives rise to an emergent strong 
U(1) symmetry in the Liouvillian.

To compare the time scales of the growth of Rényi-2 and conventional correlators, we performed TEBD simulation for a system of length $N=180$ and extracted correlation length $\xi_{R_2}(t)$ and $\xi_{\text{reg}}(t)$ from the Rényi-2 correlator $R_2(|i-j|,t)\sim\ee^{-r/\xi_{R_2}(t)}$ and $C(|i-j|,t)\sim\ee^{-r/\xi_{\text{reg}}(t)}$, respectively.
The results are shown in Fig.~\ref{fig:2}(b).
The Rényi-2 correlation length $\xi_{R_2}$ still grows exponentially fast, indicating the onset of strong-to-weak SSB at the timescale of $\sim\ln L$.
By contrast, the conventional correlation length grows only algebraically $\xi_{\text{reg}}\propto t^{z}$ with $z\simeq 0.27$.
Therefore, for a finite system, the modified model again also exhibits a intermediate time window
$\ln L\lesssim t\lesssim L^{1/z}$ in which the Rényi-2 correlator is already nonlocal while the conventional correlator remains short-ranged. 
These results further suggests that discrete symmetry $\mathbb{Z}_2$ generically leads to exponentially fast dynamical scaling, independent of the Liouvillian gap and detailed dynamical structure.

\subsection{Information-theoretic characterizations}\label{sec:info}
While correlation functions are the most direct probe of symmetry breaking orders, (strong-to-weak) spontaneous symmetry breaking can also be detected via information-theoretic quantities.
As established by recent works~\cite{Sang2025,Min2025,hauser2026strongtoweaksymmetrybreakingopen,Zhang2025,kuno2026,lu2026nonequilibriumtopologicalresponsecharge}, the conditional mutual information (CMI) can be used as a probe of mixed-state phases and phase transitions.
When the CMI between two distant regions decays exponentially with the width of an intermediate buffer, the decay length is termed as the \textit{Markov length}.
Finite Markov length ensures the local recoverability of the system~\cite{Sang2025}, while a divergent Markov length is diagnostic of mixed state phase transitions and SWSSB.
Moreover, it can be shown that the mutual information (MI) is related to the conventional correlation function~\cite{Wolf2008}, while the Rényi-2 CMI is closed related to the Rényi-2 correlation function~\cite{Min2025,Zhang2025}.
It is natural to ask whether the dynamical scaling of information-theoretic quantities, including MI and CMI, would exhibit similar behaviors as correlation functions.
In this section, we show that in $\mathbb{Z}_2$-symmetric models, the Rényi-2 Markov length extracted from CMI also grows exponentially fast in time showing the fast onset of SWSSB, while the MI grows diffusively just as the conventional correlator.

We first briefly review the basic definitions.
For the standard Markov length geometry, one takes $A$ and $C$ as two distant regions separated by a buffer $B$.
The CMI between $A$ and $C$ conditioned on $B$ is defined as
\begin{equation}\label{eq:MI}
\begin{aligned}
    I(A{:}C|B)&=I(A{:}BC)-I(A{:}B),\\
    &=S_{AB}+S_{BC}-S_{B}-S_{ABC}.
\end{aligned}
\end{equation}
where $S_{X}:=-\Tr[\rho_X\ln\rho_X]$ is the von Neumann entropy of the reduced density matrix. 
Intuitively, CMI measures the net mutual correlation that is not mediated by the buffer.
Outside critical phases, the CMI decays exponentially with the distance between $A$ and $C$ regions, namely $I(A{:}C|B)\sim\exp(-\operatorname{dist}(A,C)/\zeta)$, where $\zeta$ is defined to be the Markov length.
In the subsequent calculations, we use the Rényi-2 version of CMI:
\begin{equation}
    I_2(A{:}C|B){=}S^{(2)}_{AB}+S^{(2)}_{BC}-S^{(2)}_B-S^{(2)}_{ABC},
\end{equation}
where $S^{(2)}_X=-\log\Tr\rho_X^2$ is the Rényi-2 entropy.
The Rényi-2 Markov length $\zeta^{(2)}$ is similarly defined.
This version is naturally compatible with the Rényi-2 correlator $R_2$ and easier to handle in TEBD calculations.
Moreover, it can be written as the connected correlation function of in the double Hilbert space, thus indicative of phase transitions in the Choi states~\cite{Zhang2025,Min2025}.

\begin{figure}[t]
    \centering
    \includegraphics[width=\linewidth]{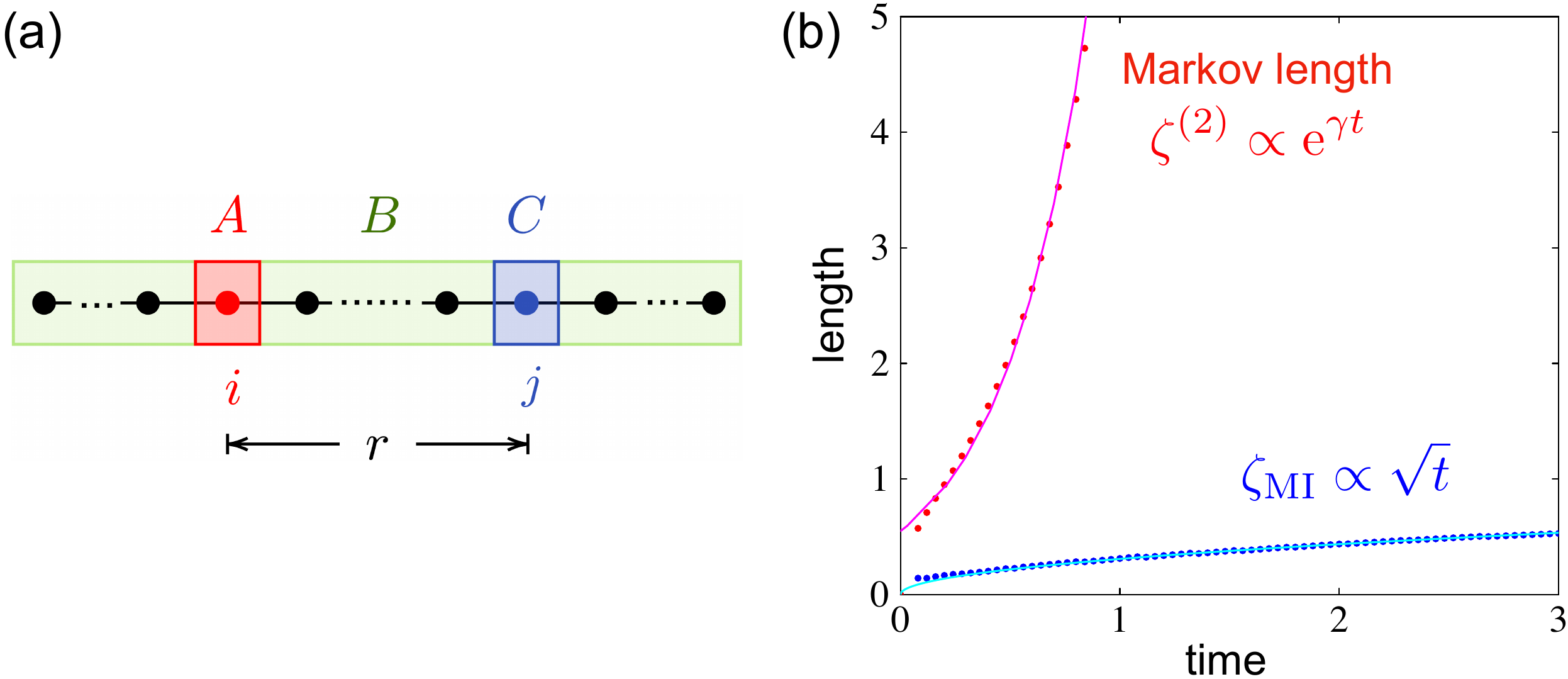}
    \caption{
    (a) Tripartition of the system into $A$, $B$, and $C$, where $B$ with a thickness of $r$ serves as the ``buffer" region in the CMI $I(A:C|B)$.
    (b) Growth of Rényi-2 Markov length $\zeta^{(2)}$ and mutual information decay length $\zeta_{\text{MI}}$, extracted from TEBD simulation of the model Eq.~\ref{eq_gaplessZ2} with $\gamma_1=\gamma_2=1$ and $L=150$. 
    }
    \label{fig:3}
\end{figure} 

In our numerical simulation, we take $A=\{i\}$, $C=\{j\}$ to be two points sitting at $i=L/3$ and $j=L/3+r$ for an open-boundary chain of length $L=150$.
The buffer $B=\overline{A\cup C}$ to taken to be the rest of the chain [Fig.~\ref{fig:3} (a)].
We use the same initial state $\ket{\psi_0}=\otimes_j\ket{\rightarrow}_j$ and calculate the mutual information $I(A{:}C)$ and Rényi-2 CMI $I_2(A{:}C|B)$ under the Lindbladian time evolution given by Eq.~\eqref{eq_gaplessZ2}.
At each time step, we scan the separation $r$ between $A$ and $C$ and extract the Rényi-2 Markov length $\zeta^{(2)}$ and mutual information decay length $\zeta_{\text{MI}}$ from
\begin{equation}
    I(A{:}C)\simeq\ee^{-r/\zeta_{\text{MI}}},\ I_2(A{:}C|B)\simeq\ee^{-r/\zeta^{(2)}}.
\end{equation}
The results are shown in Fig.~\ref{fig:3}(b).
We find that the two information-theoretic length scales separate precisely in the same way as the two correlation length scales.
The R\'enyi-2 Markov length [red dots] increases exponentially fast in time, $\zeta^{(2)}\propto \ee^{\gamma t}$, while the decay length of the ordinary mutual information correlation length [blue dots] grows diffusively, $x\propto\sqrt{t}$.
As a result, for a finite system of length $L$, there exists a broad time window, $\ln L\lesssim t\lesssim L^2$, in which the system has a long-range CMI while the mutual information stays short-ranged.
These results demonstrate that the same dynamical separation revealed by the correlation functions is also visible via information-theoretic diagnostics.
The R\'enyi-2 CMI tracks grows of the nonlinear correlation function in the double Hilbert space and inherits its ultra-fast spreading.
Meanwhile, the ordinary MI follows the slow diffusive dynamical channel associated with the conventional correlation grows.

\section{Power-law scalings in the U(1)-symmetric models}\label{sec:U1}
In this section, we discuss strongly U(1)-symmetric Lindbladians and show that they generally exhibit power-law growth of the R\'enyi-2 correlation length, in sharp contrast to the exponential growth in the $\mathbb{Z}_2$ symmetric models.
We analytically prove that in the dilute limit with one particle or one hole, the R\'enyi-2 correlation exhibits a diffusive scaling $\xi \sim t^{1/2}$, as expected by the gapless spectrum~\cite{Ogunnaike2023,Huang2025,Gu2025}.
In contrast, at any finite particle filling, due to mode coupling in the nonlinear channel, we can no longer reduce to single-particle diffusion. 
Using TEBD, our numerical simulations show that the late-time scaling follows a ballistic behavior, i.e., $\xi\propto t$. 
Such ballistic behavior remains robust even when additional U(1)-symmetric coherent perturbations are introduced, rendering the model non-integrable, as verified by our numerical simulations. 

Consider a 1D spinless-fermion chain with onsite dephasing whose Hamiltonian and jump operators are
\begin{equation}\label{eq_U1Model}
\begin{aligned}
    H&=\sum_{j=1}^{L}J(c_j^\dagger c_{j+1}+c_{j+1}^\dagger c_j)+Vn_j n_{j+1},\\
    L_j&=\sqrt{\gamma}n_j,\ j=1,2,\cdots,L.
\end{aligned}
\end{equation}
We first consider the case $V=0$, whose exact many-body Liouvillian spectrum is solvable by Bethe Ansatz~\cite{Medvedyeva2016}.
The Lindbladian of this model has strong U(1) symmetry given by the total particle number $N=\sum_j n_j$ as one can verify that $[N,H]=[N,L_j]=0$.
Hence, the double Hilbert space $\mathcal{L}(\mathcal{H})=\mathcal{H}_L\otimes \mathcal{H}_R$ can be decomposed according to the conserved U(1) charge $(N_L, N_R)$.
For an initially strongly symmetric state in the $(N_L,N_R)=(n,n)$ sector, the Lindbladian evolution
\begin{equation}
    \hat{\mathcal{L}}\rho=-\ii[H,\rho]-\frac{\gamma}{2}\sum_{j=1}^L [n_j,[n_j,\rho]]
\end{equation}
remains in the same sector.
We will consider the R\'enyi-2 correlator in Eq.~\eqref{eq_Renyi2} with $O_i=c_i$ as the local U(1) charge operator.
As will be shown later, the long-time behavior of $R_2$ crucially depends on the filling fraction $\nu=n/L$.

\begin{figure}[t]
    \centering
    \includegraphics[width=\linewidth]{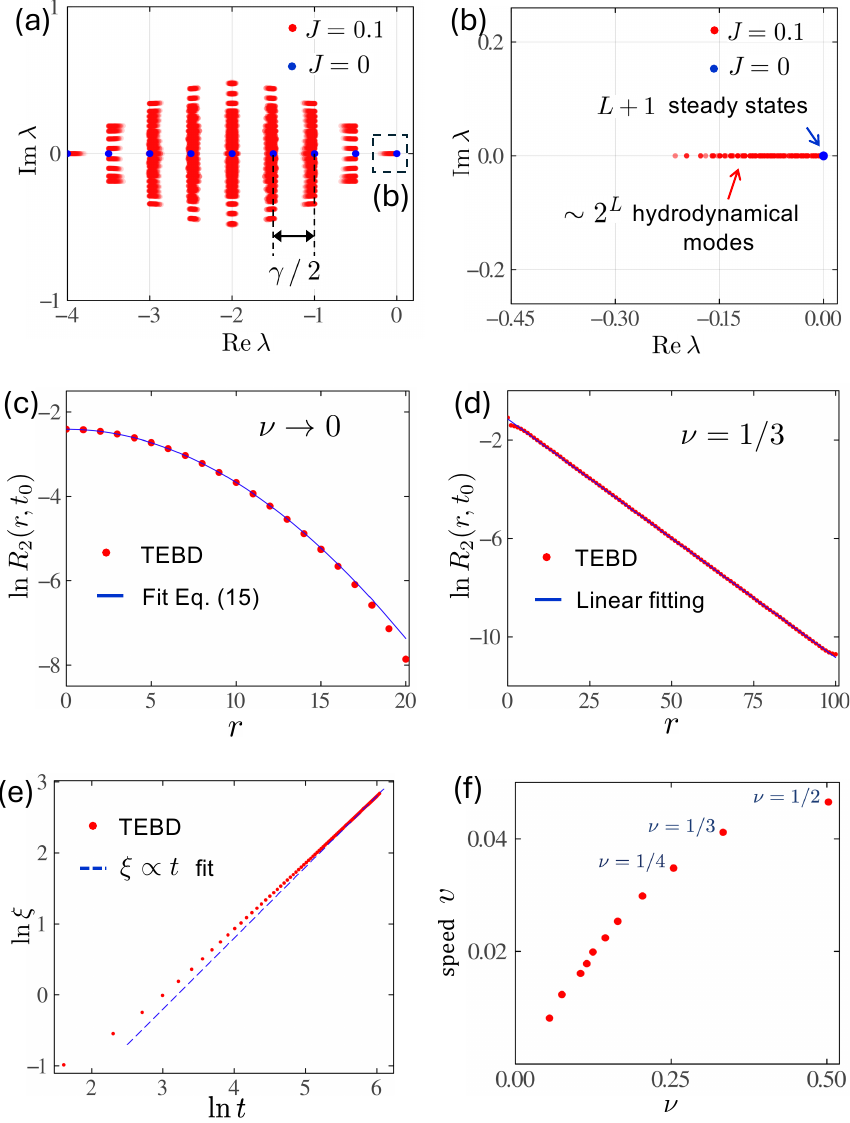}
    \caption{(a) The full Liouvillian spectrum of strongly symmetric U(1) model in Eq.~\eqref{eq_U1Model} with $\gamma=1.0$ and size $N=8$.
    In the pure dephasing limit [$J=0$], the Liouvillian eigenvalues [blue dots] are highly degenerate dots at $-\gamma n/2$ for $n=1,\cdots,N$.
    With weak coherent dynamics [$J=0.1$], the degeneracies are lifted [red dots].
    (b) Liouvillian eigenvalues [red dots] of the slow subspace near the steady state [blue dot].
    (c) When only a single particle is filled, $R_2(r,t)$ follows Eq.~\eqref{eq:single_magnonR2} [blue line], matching well with TEBD results [red dots];
    (d) At finite filling [$\nu=1/3$], $R_2(r,t)$ decays exponentially in $r$ at any $t$.
    We extract $\xi$ from the linear fitting $\ln R_2=-r/\xi$.
    (e) Log-log plot of $\xi(t)$ at $\xi=1/3$.
    At late time, the TEBD numerical results can be fitted by a linear line of slope 1 [blue dashed line], suggesting ballistic propagation.
    (f) The propagation speed $v$ of the ballistic front for different filling factors $\nu$. $v$ increases roughly following a parabolic curve peaked at $\nu=1/2$.
    The $\nu>1/2$ part is symmetric due to particle-hole symmetry.
    }
    \label{fig:4}
\end{figure} 

We first determine the slow Liouvillian spectrum and eigenmodes.
In the pure dephasing limit ($J=0$, $\gamma>0$), we use $\mathbf{n}\in\{0,1\}^L$ to denote fermion-occupation strings.
Then, every states $\kb{\mathbf{n}}{\mathbf{m}}$ in this basis is an eigenstate with eigenvalue $-\gamma |\mathbf{m}-\mathbf{n}|^2/2\in(-\gamma/2)\mathbb{Z}$.
Consequently, the spectrum consists of highly degenerate real eigenvalues equally spaced by $\gamma/2$ [blue dots in Fig.~\ref{fig:4} (a)].
Moreover, the steady-state manifold is the diagonal subspace $\Delta:=\{\kb{\mathbf{n}}{\mathbf{n}}\}$ which is $2^L$-fold degenerate.
Turning on coherent dynamics $\mathcal{L}_1:=-\ii[H,\cdot]$ lifts the degeneracy and yields gapless hydrodynamical modes [Fig.~\ref{fig:4} (b)].
To see this, we can treat $\mathcal{L}_1$ as perturbation and obtain the effective Liouvillian acting on space $\Delta$ [see Appendix.~\ref{A:SlowLin} for derivations or~\cite{Cai2013}]:
\begin{equation}\label{eq:Heisenberg}
    \mathcal{L}_{\text{eff}}=J_{\text{eff}}\sum_{j=1}^L \left(\mathbf{S}_j\cdot \mathbf{S}_{j+1}-\frac{1}{4}\right)
\end{equation}
where $J_{\text{eff}}=4J^2/\gamma>0$.
Notice that we vectorize the basis of $\Delta$ as $|\mathbf{n}\rangle\! \rangle:=\kb{\mathbf{n}}{\mathbf{n}}$ and $\mathcal{L}_{\text{eff}}$ acts on the configuration states $|\mathbf{n}\rangle\! \rangle$, which we equivalently view as a spin-1/2 chain by identifying $n_j=1(0)$ as ${\uparrow}({\downarrow})$.
It is convenient to define $H_{\text{eff}}:=-\calL_{\text{eff}}$, which maps the slow modes to the gapless excitations of a ferromagnetic Heisenberg spin chain.
Owing to the emergent SU(2) symmetry of $\calL_{\text{eff}}$, the current steady state manifold contains exactly one steady state $|\Omega_n\rangle\!\rangle$ for each filling sector $n=0,1,\cdots,L$; these are exactly the fully symmetric Dicke states~\cite{Dicke1954Coherence}.
Plugging these Dicke states into Eq.~\eqref{eq_Renyi2}, the R\'enyi-2 correlator of the steady state in each $n$-sector is given by
\begin{equation}
    R_2(r,t\to\infty)=\frac{n(L-n)}{L(L-1)},\ \forall r>0
\end{equation}
Therefore, except for the $n=0,L$ sector, the steady states has long-range R\'enyi-2 correlation. 

The closed-form solution of $R_2(r,t)$ can be obtained analytically for the single-particle (or single-hole) sector with $n=1$ (or $n=L-1$).
According to the Bethe Ansatz, the eigenmodes $\rho_q=\sum_\ell \ee^{\ii q\ell} c_\ell^\dagger\kb{\Omega_0}{\Omega_0}c_\ell$ are plane-wave excitations and the corresponding eigenvalues are $\lambda_q=-J_{\text{eff}}(1-\cos q)$.
If we take initial state as a pure state with one fermion occupying $j=0$, then, $\rho(t)=(1/L)\sum_q\ee^{\lambda_q t}\rho_q$.
Plugging into Eq.~\eqref{eq_Renyi2}, we obtain the R\'enyi-2 correlator
\begin{equation}\label{eq:single_magnonR2}
    R_2(r,t){=}{\frac{1}{L}}{\frac{{\sum_q} \ee^{\lambda_q t} }{{\sum_q} \ee^{2\lambda_q t}}}{{\sum_{p}}\ee^{\lambda_p t+\ii pr}}{=}\frac{I_0(J_{\text{eff}}t)}{I_0(2J_{\text{eff}}t)}I_r(J_{\text{eff}}t)
\end{equation}
where $I_r$ is the modified Bessel function of the first kind. 
This analytical result matches well with our TEBD numerics [Fig.~\ref{fig:4} (c)].
At late time ($J_{\text{eff}}t\gg 1$), the asymptotic form of Eq.~\eqref{eq:single_magnonR2} is given by $R_2(r,t)\simeq(\pi J_{\text{eff}}t)^{-1/2}\exp(-r^2/2J_{\text{eff}}t)$.
Therefore, the propagation of R\'enyi-2 correlation here is diffusive in the single-particle sector: it takes $t_c\propto L^2$ for the R\'enyi-2 correlation to spread across the whole system.

However, when we fill up a finite proportion of electrons ($0<\nu<1$), the nontrivial interaction between Bethe magnons yields qualitatively different dynamical scaling: The propagation of R\'enyi-2 correlation changes from diffusive to ballistic.
Due to the complexity of the scattering matrix at finite filling, obtaining a closed-form analytical expression for $R_2(r,t)$ appears intractable.
We therefore resort to TEBD simulation of the Lindbladian dynamics under Eq.~\eqref{eq_U1Model} with $\gamma=1.0, J=0.1$ and calculate the spatiotemporal profile of $R_2(r,t)$ in open-boundary chain of size $L=201$.
We numerically calculated cases with filling factors $\nu=\frac{1}{2},\frac{1}{3},\cdots,\frac{1}{10},\frac{1}{13},\frac{1}{20}$.
Due to particle-hole symmetry, the cases with filling $\nu>1/2$ are symmetric to the results with $\nu'=1-\nu$.
The initial state is chosen to be a charge density wave state $\rho_0=\kb{\psi_0}{\psi_0}$ where
\begin{equation}
    \ket{\psi_0}=\prod_{j\in\mathbb{Z}}c_{1+j/\nu}^\dagger\ket{\Omega}.
\end{equation}
The R\'enyi-2 correlator of $\rho_0$ is localized as $R_2(r,t=0)=\delta_{r,0}$, so $\rho_0$ indeed preserves the strong U(1) symmetry.

At any given time $t$, we observe that $R_2(r,t)$ follows the form $\ee^{-r/\xi(t)}$ where we can subtract $\xi(t)$ as the Rényi-2 correlation length [Fig.~\ref{fig:4} (d)], in direct contrast to the single-particle case shown in Fig.~\ref{fig:4} (c).
We numerically obtained $\xi(t)$ for $t\in(0,425)$, where we plot the result in log-log scale [$\nu=1/3$ case is shown in Fig.~\ref{fig:4} (e) as an example].
At late time, the $\ln\xi$ curve becomes almost linear and the slope $\dd \ln \xi/\dd \ln t$ is trending to 1 [blue dashed line in Fig.~\ref{fig:4} (e)], suggesting ballistic dynamical scaling.
The actually slope in the numerical data is slightly below 1: by fitting data in the range $t\in(400,425)$, we get slope $0.967\simeq 1$.
This deviation could potentially be attributed to the truncation error in TEBD evolution.
We also tested the robustness of the ballistic scaling law in non-integrable models by turning on $V\neq0$ and also away from the $J\ll\gamma$ limit [see Appendix.~\ref{A:Stability}].

Next, we extract the propagation speed $v$ of ballistic front by fitting the slope of the $\xi(t)$ graph.
First of all, we find that in the strong dissipation limit $\gamma\gg J$, the speed $v$ is proportional to the effective coupling $J_{\text{eff}}=4J^2/\gamma$.
This strongly suggests the ballistic speed is emergent infrared phenomenon of the effective Liouvillian in the hydrodynamical sector.
Furthermore, we extracted the ballistic speed for filling factors $\nu\in[\frac{1}{20},\frac{1}{2}]$ as shown in Fig.~\ref{fig:4} (f). The $\nu>1/2$ part is symmetric, forming a dome-shape curve which can be phenomenologically captured by $\nu(1-\nu)$.
As $\nu$ increases from the dilute limit, the ballistic speed first increases linearly in $\nu$, and then saturates at $\nu\simeq 1/2$.
This suggests the ballistic channel is the strongest when the system has the largest dynamical phase space ($\nu=1/2$).
However, when $\nu$ approaches $0$ or $1$, the ballistic speed decreases linearly to zero, eventually reducing to the single-particle case with diffusive scaling.

\section{Conclusion and discussion}\label{sec:outlook}
In this work, we investigated the late-time dynamical scaling of SWSSB in one-dimensional open quantum systems. 
Our main conclusion is that the asymptotic spreading behavior of the Rényi-2 correlator is not determined by whether the Liouvillian spectrum is gapped or gapless. 
Instead, the key ingredient is the interplay between symmetry and the dynamical sectors that actually contribute to the nonlinear correlator. 
This leads to two sharply distinct classes of late-time behavior. 
For strongly $\mathbb{Z}_2$-symmetric Lindbladians, the Rényi-2 correlation length grows exponentially in time, $\xi(t)\sim \ee^{\lambda t}$, even when the Liouvillian spectrum is gapless. 
By contrast, for strongly symmetric $U(1)$ Lindbladians, the growth is ballistic with $\xi(t)\sim t$ at finite filling but becomes diffusive at zero-filling limit. 
These symmetry-enforced late-time dynamical scalings are universal and remain robust even when integrability is broken by coherent perturbations, as verified by our numerical simulations. 

Several open directions naturally follow from this work. 
Very recently, similar ballistic behavior has been observed in U(1) symmetric decohered spin-1/2 chain and rotor models, where the ballistic scaling is analytically obtained for the decohered rotor model~\cite{hauser2026strongtoweaksymmetrybreakingopen}.
Hence, it would be interesting to develop a coarse-grained theory for late-time scaling behavior of Rényi-$n$ correlator in U(1) systems with finite fillings. 
Second, our analysis raises the broader question of how symmetry constrains late-time dynamical scaling in other symmetry classes, including non-Abelian symmetries and higher-form symmetries~\cite{song2025strongtoweakspontaneoussymmetrybreaking,ziereis2025strongtoweaksymmetrybreakingphases,Zhang2025}.
Third, our work focuses on one-dimensional systems, where the SWSSB transition occurs only at infinite time in the thermodynamic limit, making the late-time dynamical scaling with system size particularly important. In higher dimensions ($d>1$), however, it remains unclear how the Liouvillian gap and symmetry structure influence the dynamics of the Rényi-$n$ correlators when the infinite-time SWSSB occurs.
Finally, the examples discussed in this paper are experimentally accessible in programmable open quantum platforms including trapped Rydberg ions~\cite{Barreiro2011,Noel2022MIPTTrappedIon}, optical lattices~\cite{OpticalLatticeReview}, and neutral atom arrays~\cite{Evered2023NeutralAtomCZ,Lis2023MidcircuitNeutralAtoms}. 
Our $\mathbb{Z}_2$-symmetric model can be realized in a quantum circuit~\cite{Barreiro2011,Andersen2019,Hoke2023,Shah2025} by implementing CNOT gates between nearest-neighbor lattice sites. 
The U(1)-symmetric model corresponds to coherent hopping in the presence of onsite dephasing, a setting that is readily realizable in experiments on dissipative quantum systems.
The nonlinear correlator characterizing the SWSSB are typically two-copy quantities, whose detection requires doubled-copy protocols~\cite{Daley2012} or trajectory-based reconstructions~\cite{Naghiloo2016,Flurin2020}. There also has been multiple proposals and active discussion for the detection of SWSSB in experiments~\cite{FengLei2025PRL,Liu2025,Weinstein2025,Feng2025}.
Our results also provide direct insights into the preparation of SWSSB steady states in experiments with decoherence channels.

\section{Acknowledgement}
We thank Ceren B. Dağ and Cenke Xu for helpful discussions.
CS, KZ, and KS was supported by the Office of Naval Research (Grant No. MURI N00014-20-1-2479) and the National Science Foundation through the Materials Research Science and Engineering Center at the University of Michigan (Award No. DMR-2309029). YY
acknowledges support from NSF under award number
DMR-2439118.
\\
\vspace{1em}

\appendix

\section{\texorpdfstring{$\mathbb{Z}_2$}{Z2} symmetric gapped models: Exact solution in the Pauli-string basis and exponentially fast spreading}\label{A:Z2Gap}
In this section, we show a typical $\mathbb{Z}_2$ model with gapped Liouvillian spectrum, leading to the exponentially fast spreading of the R\'enyi-2 correlation.
Specifically, we consider a one-dimensional chain of length $L$ with a qubit on each site, and the jump operators are given by
\begin{equation}
  L_j=\sqrt{\gamma}\, Z_j Z_{j+1},\qquad j=1,\ldots,L-1.
\end{equation}
The system has strong $\mathbb{Z}_2$ symmetry, where the symmetry operator is given by $X=\prod_{j=1}^L X_j$ and we have $[H,X]=[L_j,X]=0$ for any $j=1,\cdots,L$.

This model becomes exactly solvable in the Zeno limit where the coherent dynamics is turned off [$H=0$]. Now, the evolution of $\rho(t)$ is given by
\begin{equation}
  \partial_t\rho=\hat{\cal L}\rho
  =\gamma\sum_{j=1}^{L-1}\Big(Z_jZ_{j+1}\rho Z_jZ_{j+1}-\rho\Big).
  \label{eq:Zeno_Lindblad}
\end{equation}
In this limit, the model can be directly mapped to the quantum channel $\mathcal{E}=\prod_{\langle ij\rangle}\mathcal{E}_{ij}$ with $\mathcal{E}_{ij}[\rho]=pZ_i Z_j\rho Z_j Z_i+(1-p)\rho$ where $p=(1-\ee^{-\gamma t})/2$.~\cite{Lessa2025PRXQ,Guo2025}
Physically, each jump $L_j\propto Z_jZ_{j+1}$ continuously monitors the bond $X$-parity.
As a result, operator components that anticommute with a monitored bond are exponentially suppressed, while commuting components survive.
This is most transparent in the Pauli-$X$ string basis.
For any subset $U\subseteq\{1,\dots,L\}$, define the normalized string
$P_U=2^{-L/2}\prod_{j\in U}X_j$ (with $P_\emptyset=I/2^{L/2}$), which satisfies $\Tr(P_U^\dagger P_{U'})=\delta_{U,U'}$.
Using $ZXZ=-X$, conjugation by $Z_jZ_{j+1}$ flips the sign of $P_U$ if the string has $X$ on exactly one endpoint of the bond $(j,j+1)$.
Equivalently, only bonds where the pattern switches between $I$ and $X$ contribute to the incoherent decay.
Let $n_{\rm dw}(U)$ count these ``domain walls'', then $\hat{\cal L}$ is diagonal in this Pauli-$X$ string basis $\hat{\cal L}P_U=-2\gamma\,n_{\rm dw}(U)P_U$ and we can write the time evolution of the density matrix as
\begin{equation}\label{eq:PU_diag_and_solution}
  \rho(t)=\sum_U c_U\,e^{-2\gamma t\,n_{\rm dw}(U)}P_U
\end{equation}
Thus the dissipation penalizes spatial variation of operator strings: each domain wall costs a decay rate $2\gamma$.
At long times only $n_{\rm dw}(U)=0$ survives, i.e., the uniform strings $I$ and $X$ (the maximally mixed state within the corresponding strong-symmetry sector).

To study the scrambling dynamics of Rényi-2 correlator, we take the strongly symmetric product state
$\rho(0)=\bigotimes_j\ket{\rightarrow}\!\bra{\rightarrow}_j=2^{-L/2}\sum_U P_U$
and consider the Rényi-2 correlator
$R_Z(i,j;t)\equiv \Tr[\rho(t)Z_iZ_j\rho(t)Z_jZ_i]/\Tr[\rho(t)^2]$.
Using the orthonormality, we have
\begin{equation}
  R_Z(i,j;t)=
  \frac{\sum_U s_i(U)s_j(U)\,e^{-4\gamma t\,n_{\rm dw}(U)}}
  {\sum_U e^{-4\gamma t\,n_{\rm dw}(U)}}.
  \label{eq:RZ_sum}
\end{equation}
This is naturally interpreted as a classical ferromagnetic Ising correlator: the variables $s_j(U)$ play the role of Ising spins, while $n_{\rm dw}(U)$ counts anti-parallel bonds, so the weight is proportional to
$\exp\!\big(2\gamma t\sum_{j=1}^{L-1}s_js_{j+1}\big)$.
In particular, in 1D, one obtains $R_Z(i,j;t)\simeq(\tanh(2\gamma t))^{|i-j|}=e^{-|i-j|/\xi(t)}$ with
$\xi(t)=1/[-\ln\tanh(2\gamma t)]\simeq e^{4\gamma t}/2$ for $\gamma t\gg1$~\cite{Sachdev2011QPT}.
Hence correlations spread across a finite chain on the short timescale $t_f\sim (4\gamma)^{-1}\ln L$.

\section{Proving the diffusion-reaction equation leads to gapless Liouvillian spectrum}\label{A:Z2Gapless}
This Appendix section is dedicated to proving that the Markov equation Eq.~\eqref{eq:MarkovChain} implies a gapless Liouvillian spectrum, with finite-size scaling as $\delta\propto L^{-2}$.

First of all, notice that a diagonal state $\kb{\mathbf{z}}{\mathbf{z}}$ remains diagonal during the evolution.
Thus, the diagonal subspace $\Xi^+$ is an invariant subspace closed under the Lindbladian evolution.
The Lindbladian is therefore block diagonal and denote $\mathcal{L}_{\text{diag}}$ as the block acting on the diagonal subspace.
Since the spectrum of $\mathcal{L}_{\text{diag}}$ is a subset of the entire spectrum of the system $\sigma(\mathcal{L}_{\text{diag}})\subset\sigma(\mathcal{L})$, if the spectrum of this subspace is gapless, then the whole system is gapless.
In the following we prove that the action of $\mathcal{L}_{\text{diag}}$ indeed leads to a gapless Liouvillian spectrum.
Due to the action of $L^{(2)}$, the number of $\mathbb{Z}_2$ domain walls, denoted as $n_{\text{dw}}$, is decreasing monotonically during the evolution.
When taking periodic boundary conditions, the number of $\mathbb{Z}_2$ domain walls is even.
We arrange the basis according to the domain-wall numbers, and in such a basis, the $\mathcal{L}_{\text{diag}}$ is strictly block lower-triangular in this form:
\begin{equation}\label{eq:blockdia}
    \mathcal{L}_{\text{diag}}=\mqty(
    \ddots & \vdots & \vdots & \vdots \\
    \cdots & \mathcal{L}_4 & 0 & 0 \\
    \cdots & * & \mathcal{L}_2 & 0 \\
    \cdots & * & * & \mathcal{L}_0
    )
\end{equation}
and the spectrum satisfies $\sigma(\mathcal{L}_{\text{diag}})=\sigma(\mathcal{L}_0)\cup\sigma(\mathcal{L}_2)\cup\cdots$ where $\mathcal{L}_{n}$ denotes the Liouvillian block acting on the $n$-domain wall subspace.
$\mathcal{L}_0$ has zero eigenvalue since it belongs to the steady state space.
Therefore, it suffice to show that the $\mathcal{L}_2$ spectrum is gapless.

We can label a two-domain-wall state by their coordinates $(r_1, r_2)$.
As we take periodic boundary conditions for the system, the coordinates $r_1$ and $r_2$ are all modulo system size $L$ and satisfy $r_1\neq r_2$.
The configuration space $\mathcal{M}:=\{(r_1,r_2)\sim (r_2, r_1)|r_i\in S^1,r_1\neq r_2\}$ is in fact a M\"obius strip.
For now, we neglect the global boundary condition of this configuration space where $(r_1, r_2)$ live, and just focus on a local coordinate patch where $|r_1-r_2|>1$.
Using $|r_1, r_2\rangle\!\rangle$ to denote a two-domain state, via Eq.~\eqref{eq:MarkovChain}, we see that
\begin{equation}
\begin{aligned}
    \calL_2|r_1, r_2\rangle\!\rangle{=}&\gamma_1|r_1{+}1, r_2\rangle\!\rangle
    {+}\gamma_1|r_1, r_2{+}1\rangle\!\rangle{+}\gamma_1|r_1{-}1, r_2\rangle\!\rangle\\
    &{+}\gamma_1|r_1, r_2{+}1\rangle\!\rangle{-}4\gamma_1|r_1, r_2\rangle\!\rangle
\end{aligned}
\end{equation}
Therefore, the bulk of the configuration space can be viewed as a tight-binding model on square lattice with hopping amplitude towards neighboring site given by $-\gamma_1$.
Therefore, if we modify the boundary condition such that the configuration space $\mathcal{M}$ is a torus, we can use Bloch theorem and define $k_1$ and $k_2$ and the spectrum would be given by
\begin{equation}
    \lambda_2(k_1,k_2)=2\gamma_1(\cos k_1 +\cos k_2-2)
\end{equation}
Therefore, the spectrum obtained via assuming a torus-shaped $\mathcal{M}$ is gapless, whose low-energy excitations are diffusive $\lambda_2(\delta\mathbf{k})\sim-\gamma_1|\delta\mathbf{k}|^2$. 
Here, we notice that $\mathcal{L}_2$ is Hermitian, hence there is no Liouvillian skin effect~\cite{Haga2021,Feng2024}, i.e., spectral sensitivity to the modification of boundary conditions.
In other words, modifying the boundary condition can at most alter $\sim \mathcal{O}(L)$ states out of $\sim \mathcal{O}(L^2)$ states in its spectrum.
Therefore, the majority of the bulk spectrum remains gapless, regardless of boundary conditions.
So, we proved that the spectrum of $\mathcal{L}_2$ is gapless, hence, the $\mathbb{Z}_2$-symmetric model we presented in the main text is gapless with $\propto L^{-2}$ finite-size Liouvillian gap scaling, consistent with the numerical observation in Fig.~\ref{fig:1} (b).

\vspace{2em}
\section{Emergent U(1) symmetry and t' Hooft anomaly in the low energy sector}\label{app:U1}

Restricting to the low energy part of the Liouvillian spectrum, the projected Lindbladian takes the block-diagonal form shown in Eq.~\ref{eq:blockdia}. The gapless modes arise from the diagonal blocks,
\[
\sigma(\mathcal{L}_{\mathrm{diag}})=\sigma(\mathcal{L}_0)\cup\sigma(\mathcal{L}_2)\cup\cdots,
\]
where $\mathcal{L}_n$ denotes the sector with $n$ domain walls. This block-diagonal structure reveals an \textit{emergent strong $U^{\operatorname{dw}}(1)$ symmetry} in the low-energy sector, corresponding to conservation of domain-wall number generated by~\footnote{This $U(1)$ symmetry is not onsite.}:
\[
U^{\text{dw}}(\theta)=\ee^{\ii\theta\sum_i Z_i Z_{i+1}}.
\]
 This symmetry is absent in the full Lindbladian, but emerges after projection onto the gapless sector.

The origin of this emergent symmetry is transparent from the jump operators. The term $L^{(1)}_j= X_j P_{j-1,j+1}$ that is responsible for the gapless spectrum preserves the total number of domain walls, since it only shifts a single domain wall in space, and therefore commutes with $U^{\text{dw}}(\theta)$. By contrast, $L^{(2)}_j=X_j P_{j,j-1}P_{j,j+1}$ annihilates a neighboring pair of domain walls and thus breaks $U^{\text{dw}}(1)$. However, this pair-decimation process contributes only to the lower-triangular part of Eq.~\ref{eq:blockdia}, and is absent after projecting onto the diagonal sector $\sigma(\mathcal{L}_{\mathrm{diag}})$. As a result, the diagonal blocks exhibit a strong \(U^{\text{dw}}(1)\) symmetry, which in turn enforces gaplessness of \(\sigma(\mathcal{L}_{\mathrm{diag}})\) and leads to long mixing times. Therefore, the gapless modes and diffusive dynamics in Sec.~\ref{sec:STNSSB} can be traced to the emergent \(U(1)\) symmetry in the low-energy sector.  

Notably, this emergent \(U^{\text{dw}}(1)\) symmetry arises whenever the jump operators governing the Markov dynamics do not increase the domain-wall number. This is precisely the situation in Secs.~\ref{sec:STNSSB}--\ref{sec:robust}, where the jump operators either move domain walls in space or annihilate a pair of domain walls. Moreover, because the Lindbladian considered in Secs.~\ref{sec:STNSSB}--\ref{sec:robust} preserves the weak symmetry \(U^{\mathrm{dw}}(1)\), it is sufficient to restrict attention to superoperators of the form \(\ket{\{z\}}\bra{\{z'\}}\), where \(\{z\}\) and \(\{z'\}\) carry the same domain-wall number. Accordingly, a Liouvillian eigenoperator can be decomposed as
\[
\rho = \sum_n \lambda_n \rho_n,
\]
where each \(\rho_n\) lies in the sector with fixed domain-wall number \(n\). In this basis, the Liouvillian superoperator takes a block lower-triangular form: 
\begin{align}\label{eq:pm}
L=
\begin{pmatrix}
L_{mm} & 0 & 0 & \cdots & 0\\
L_{m-2,m} & L_{m-2,m-2} & 0 & \cdots & 0\\
L_{m-4,m} & L_{m-4,m-2} & L_{m-4,m-4} & \cdots & 0\\
\vdots & \vdots & \vdots & \ddots & \vdots\\
L_{0m} & L_{0,m-2} & L_{0,m-4} & \cdots & L_{00}
\end{pmatrix},
\end{align}
\(L_{nn}\) acts within the subspace with fixed domain-wall number \(n\), while \(L_{mn}\) with \(m<n\) reduces the \(n\)-domain-wall sector to the \(m\)-domain-wall sector. Since the Lindbladian dynamics never increases the domain-wall number, the full Liouvillian necessarily takes a lower-triangular form.

For a lower-triangular matrix, the spectrum is completely determined by its diagonal blocks:
\[
\mathrm{spec}(L)=\bigcup_{k=0}^{m/2}\mathrm{spec}(L_{2k,2k}).
\]
Therefore, the Liouvillian spectrum of Eq.~\ref{eq:pm} coincides with that of its diagonal sector,
\[
\bigcup_{k=0}^{m/2}L_{2k,2k}.
\]
Since each diagonal block preserves the domain-wall number, this sector has a strong \(U(1)\) symmetry, which in turn guarantees that the spectrum is gapless.

In addition, the diagonal sector $\bigcup_{k=0}^{m/2}L_{2k,2k}$ carries a mixed ’t Hooft anomaly between two strong $\mathbb{Z}_2$ symmetries,
\begin{align}
    Z_2^a &:~\prod_i X_i, \\
    Z_2^b &:~\ee^{\ii\frac{\pi}{4}\sum_i Z_i Z_{i+1}}.
\end{align}
Here, $Z_2^a$ is the global $X$-parity symmetry, while $Z_2^b$ is a non-onsite CZ-type symmetry, corresponding to the subgroup of the emergent $U^{\text{dw}}(1)$ symmetry that measures the domain-wall number modulo $4$. It is worth emphasizing that $Z_2^a$ is already present at the microscopic level, since all jump operators commute with the $X$-parity operator. By contrast, $Z_2^b$ is not a symmetry of the Lindbladian, but emerges only after restricting to the diagonal sector. Because the two strong symmetries $Z_2^a$ and $Z_2^b$ carry a mixed ’t Hooft anomaly, their steady state need to be highly entangled. This follows from the theorem of Ref.~\cite{lessa2025mixed}, which states that a 1d mixed state with strong symmetries carrying a ’t Hooft anomaly is necessarily not tripartite separable. In the present case, this immediately rules out naïve strong-to-weak SSB states that are classically separable (such as $I+X$) \cite{chen2024symmetry,li2025highly,lu2025holographic}. 
Instead, the anomaly is compatible with a steady state exhibiting strong-to-nothing symmetry breaking with a GHZ-like structure.

\vspace{2em}
\section{Restoring the strong symmetry}\label{app:kw}

In the foregoing discussion, we showed that long-time Lindblad evolution can break a strong symmetry, either reducing it to a weak symmetry or destroying it altogether. In Sec.~\ref{sec:STNSSB}, the jump operator in Eq.~\ref{eq_gaplessZ2} drives the strongly symmetric product state $\otimes_i |+\rangle_i$ into a GHZ state, thereby realizing strong-to-nothing symmetry breaking. A natural complementary question is whether dissipation can instead restore symmetry. In other words, we ask whether Lindblad evolution can restore a strong symmetry when the initial state spontaneously breaks it. 

This question can be answered straightforwardly using the Kramers--Wannier duality. We therefore apply the Kramers--Wannier duality to the jump operators in Eq.~\ref{eq_gaplessZ2}.
\begin{equation}\label{eq:dual}
  \begin{aligned}
    L^{(1)}_{j} &= \sqrt{\gamma_1}\, Z_j Z_{j+1}\, (1-X_jX_{j+1}),\\
    L^{(2)}_{j} &= \sqrt{\gamma_2}\, Z_j Z_{j+1}\, (1-X_j)(1-X_{j+1}),
  \end{aligned}
\end{equation}
The first operator hops a single flipped spin \( |-\rangle_j \) to a neighboring site, while the second operator annihilates a neighboring pair of flipped spins by mapping \( |-\rangle_j |-\rangle_{j+1} \) to \( |+\rangle_j |+\rangle_{j+1} \). Throughout this dissipative process, the strong \( Z_2 \) symmetry \( \prod_i X_i \) is preserved.
If we start from the symmetry-breaking GHZ state (\( \otimes_j \ket{\uparrow}_j + \otimes_j \ket{\downarrow}_j \)) (even charge sector) and evolve it under the Lindblad dynamics in Eq.~\ref{eq:dual}, the system relaxes to the fully symmetric state \( \otimes_i |+\rangle_i \), in which the strong symmetry is completely restored. Likewise, starting from the odd charge sector,
the steady state becomes the one-particle Dicke state
$ \frac{1}{\sqrt{L}} \sum_{i=1}^L \ket{-}_i \bigotimes_{j \neq i} \ket{+}_j$
which exhibits long-range entanglement.

Notably, by applying various duality or non-invertible symmetry transformations, one can generate a whole zoology of intriguing dissipative dynamics and steady states. For example, if we apply the Jordan-Wigner transformation to the jump operators in Eq.~\ref{eq_gaplessZ2}, the resulting jump operators remain local. Starting from an atomic insulator with one fermion per site and evolving under the corresponding Lindblad dynamics, the system can be driven into a steady state realizing a p-wave topological superconductor. Likewise, if we apply the Kennedy-Tasaki transformation to the jump operators in Eq.~\ref{eq_gaplessZ2}, the resulting steady state becomes the ZXZ cluster state, a symmetry-protected topological state that can serve as a resource for measurement-based quantum computation.

\vspace{2em}
\section{Time-Evolving Block Decimation (TEBD) for Lindbladian evolutions}\label{A:TEBD}

In this section, we brief introduce how we numerically enact the Lindbladian dynamics using time-evolving blocking decimation (TEBD) and calculate the Rényi-2 correlator at each time step.

\begin{figure}[b]
    \centering
    \includegraphics[width=\linewidth]{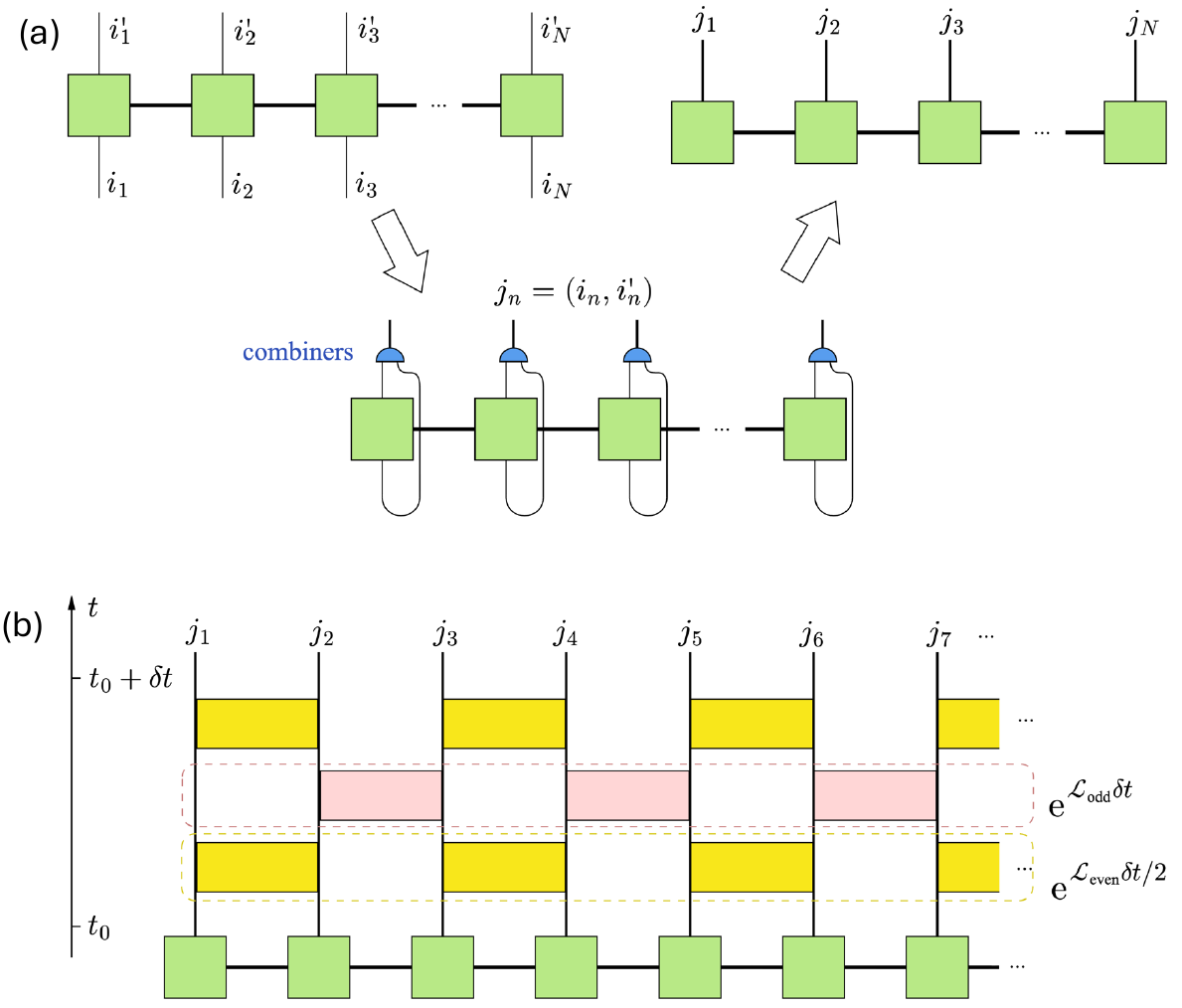}
    \caption{(a) Implementation of the Choi-Jamiołkowski isomorphism using the combiner tensors [marked in blue].
    (b) Sketch of a Trotter step.
    The evolution operator during a time step $\delta t$ is decomposed into even and odd layers, where each layer is composed of local gates which can be sequentially applied to the MPS.
    }
    \label{fig:r1}
\end{figure}

First, to perform the evolution $|\rho(t)\rangle\!\rangle=\exp(\calL t)|\rho\rangle\!\rangle$ using TEBD, we must write the density matrix into matrix-product states (MPS).
We first use Choi–Jamiołkowski isomorphism to write the density matrix as a vector in the doubled hilbert space.~\cite{Jamiolkowski1972,Choi1975}
We implement this using the combiner tensors to fuse the tensor indices in the original ``ket" and ``bra" spaces.
This process can be sketched by Fig.~\ref{fig:r1} (a).
Next, we write the superoperator $\mathcal{L}$ using the following tensor product form:
\begin{equation}
\begin{aligned}
    \calL=&-\ii(H\otimes I-I\otimes H^T)+\sum_{\mu}(L_\mu\otimes L_\mu^*\\
    &-\frac{1}{2}L_\mu^\dagger L_\mu\otimes I-\frac{1}{2}I\otimes L_\mu^T L_\mu^*).
\end{aligned}
\end{equation}
For nearest-neighbor coupled Lindbladian, $\calL$ can be break into even and odd terms as $\calL=\calL_{\text{even}}+\calL_{\text{odd}}$ where each of them is the sum of local two-site terms $\calL_{\text{even(odd)}}=\sum_{n\in\text{even(odd)}}\calL_{n,n+1}$.
We evolve the state $|\rho\rangle\!\rangle$ by continuous acting $\delta U=\exp(\calL\delta t)$ on the MPS.
To perform the time evolution, we use the second-order Trotter-Suzuki decomposition~\cite{Suzuki1990}
\begin{equation}
    \ee^{\calL\delta t}=\ee^{\calL_{\text{even}}\delta t/2}\ee^{\calL_{\text{odd}}\delta t}\ee^{\calL_{\text{even}}\delta t/2}+\mathcal{O}(\delta t^3)
\end{equation}
where $\ee^{\calL_{\text{even(odd)}}\delta t}$ can be written as joint pieces of local gates.
Each local gate is applied sequentially to the MPS, followed by singular-value decomposition and truncation controlled by cutoff dimension $\chi$ and singular value tolerance $\epsilon$ [Fig.~\ref{fig:r1} (b)].
In this way, the bond dimension remains under control throughout the non-unitary time evolution.

The Rényi-2 observables can also be naturally evaluated in the doubled Hilbert space. 
Using the Hilbert--Schmidt inner product, the R\'enyi-2 correlator can be written as
\begin{equation}
    R_2(i-j,t)=\frac{\langle\!\langle\rho(t)|(O_i\otimes O_i^*)(O_j\otimes O_j^*)|\rho(t)\rangle\!\rangle}{\langle\!\langle\rho(t)|\rho(t)\rangle\!\rangle}.
\end{equation}
Therefore, we can simply use the MPS contraction to evaluate the R\'enyi-2 correlator at each time step of the evolution.

\section{Derivation of the effective slow Lindbladian}\label{A:SlowLin}
In this section, we derive the expression for $\calL_{\text{eff}}$ as presented in Eq.~\eqref{eq:Heisenberg}.
We use $P$ to denote the projection operator into the diagonal subspace and denote $Q=1-P$.
Then, $\calL_{\text{eff}}$ acts on the slow subspace as $\dot{P\rho}=\calL_{\text{eff}}[P\rho]$.
Decompose $\rho=P\rho+Q\rho$, we get
\begin{equation}
\begin{aligned}
    \dot{P\rho}&=P\calL_1 Q\rho,\\
    \dot{Q\rho}&=Q\calL_0 Q\rho +Q\calL_1 P\rho + Q\calL_1 Q\rho.
\end{aligned}
\end{equation}
Here we have already simplified the equations using the fact $P\calL_0 P=0$.
In the regime $\gamma\gg J$, $P\rho$ evolves slowly with rate $\sim|\!|\calL_1|\!|$ while $Q\rho$ relaxes quickly with rate $\sim|\!|\calL_0|\!|$.
We then perform adiabatic elimination by taking $\dot{Q\rho}=0$ and get
\begin{equation}
    \dot{P\rho}=\left[-P\calL_1(Q\calL_0 Q)^{-1}\calL_1 P+\mathcal{O}\left(\frac{J^3}{\gamma^2}\right)\right]
\end{equation}
Therefore, to the leading order, the effective slow generator on the diagonal sector is
\begin{equation}
    \calL_{\text{eff}}=-P\calL_1(Q\calL_0 Q)^{-1}\calL_1 P
\end{equation}
Next, we write down the explicit form for the effective Liouvillian.
First of all, since $\calL_0=-(\gamma/ 2)\sum_j[n_j,[n_j,\cdot]]$ is diagonal in the diagonal sector, we have $(Q\calL_0 Q)^{-1}=-I/\gamma$. Hence,
\begin{equation}
    \calL_{\text{eff}}=\frac{1}{\gamma}P\calL_1^2 P=-\frac{1}{\gamma}P[H,[H,\cdot]]P.
\end{equation}
Next, we write the action of this reduced form of $\calL_{\text{eff}}$ on the basis state $\ket{\mathbf{n}}\!\bra{\mathbf{n}}$ where $\mathbf{n}\in\{0,1\}^L$ is the fermion-occupation string.
Denote $|\mathbf{n}\rangle\!\rangle:=\kb{\mathbf{n}}{\mathbf{n}}$, we have
\begin{equation}
    \calL_{\text{eff}}|\mathbf{n}\rangle\!\rangle=\frac{2J^2}{\gamma}{\sum_{j=1}^N}[\operatorname{SWAP}_{j,j+1}-\mathbb{I}_{j,j+1}]|\mathbf{n}\rangle\!\rangle
\end{equation}
where the $\operatorname{SWAP}_{j,j+1}$ operator swaps the qubits at site $j$ and $j+1$.
Next, we utilize the identity $\operatorname{SWAP}_{j,j+1}=(\mathbb{I}_{j,j+1}+\boldsymbol{\sigma}_j\cdot\boldsymbol{\sigma}_{j+1})/2$ and obtain
\begin{equation}
    \calL_{\text{eff}}=J_{\text{eff}}\sum_{j}\left(\mathbf{S}_j\cdot\mathbf{S}_{j+1}-\frac{1}{4}\right).
\end{equation}
\\

\section{Robustness of the linear scaling law in U(1) symmetric models}\label{A:Stability}

\begin{figure}[b]
    \centering
    \includegraphics[width=\linewidth]{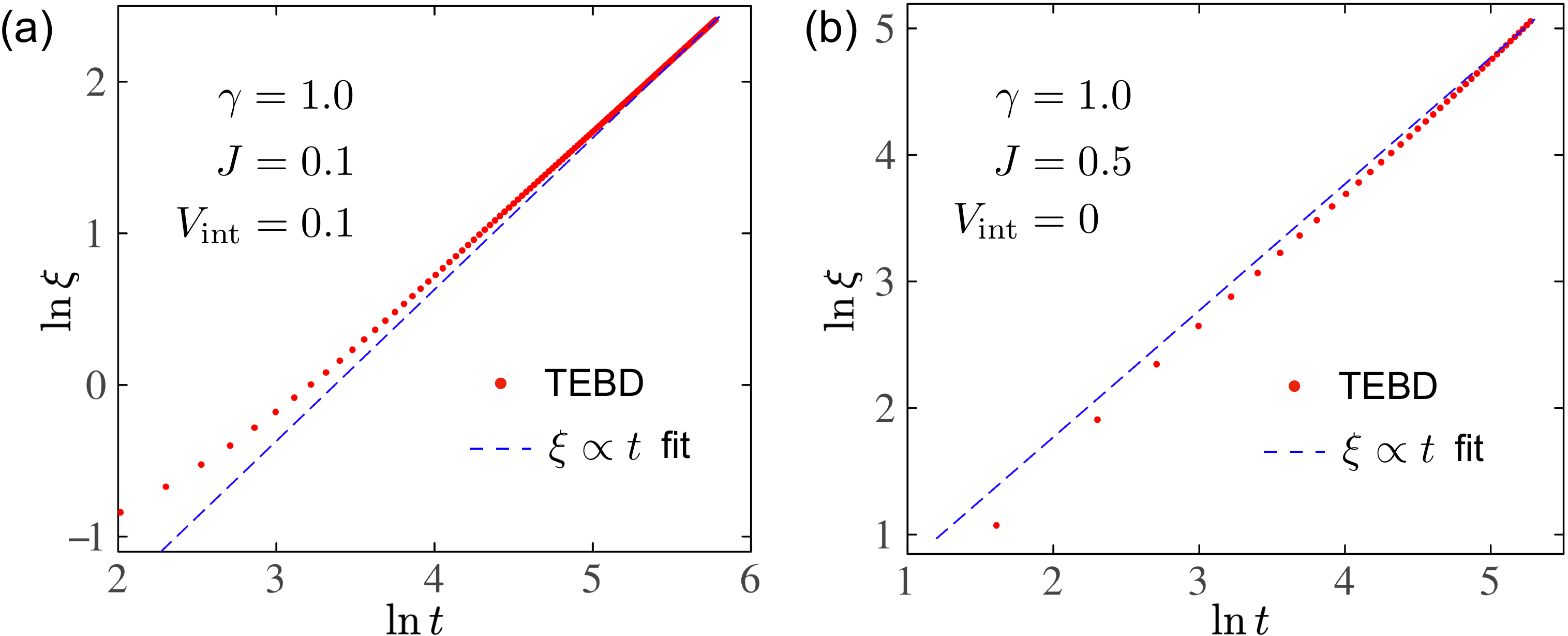}
    \caption{Universality of the ballistic dynamical scaling. (a) $\ln\xi$-$\ln t$ plot when the interaction is turned on [$L=201,J=0.1,\gamma=1.0,V=0.1$]. As the unity-slope fit suggests, the dynamical scaling is ballistic. (b) $\ln\xi$-$\ln t$ plot when $\gamma$ and $J$ are comparable with each other [$L=101,J=0.5,\gamma=0.1,V=0$].  }
    \label{fig:r2}
\end{figure}

In this section, we demonstrate the stability of the ballistic propagation in the U(1) symmetric model.
The U(1) model calculation presented in the main text is integrable when $V=0$.
Also, the effective Liouvillian in Eq.~\eqref{eq:Heisenberg} is only valid under the strong dissipative limit $J\ll \gamma$.
Here, we demonstrate examples showing that the ballistic behavior survives when the model is non-integrable and when $J\sim\gamma$.
First, we turn on the next-nearest neighbor density-density interaction in Eq.~\eqref{eq_U1Model} to make the system non-integrable.
An example is presented in Fig.~\ref{fig:r2} (a) with $\gamma=1.0,J=0.1,V=0.1$ and $\nu=1/4$.
As the log-log plot of $\xi(t)$ suggests, the ballistic dynamical scaling survives under the introduction of non-integrability.
Moreover, when we move away from the $J\ll\gamma$ limit and take $J=0.5, \gamma=0.5$, and filling factor $\nu=1/5$ in Fig.~\ref{fig:r2} (b), the ballistic propagation of the Rényi-2 correlation is still robust.

\bibliography{refs.bib}

\end{document}